\documentstyle[epsfig,psfig,11pt,supertabular]{article}

\def\and{\char'046}

\font\sl=cmsl10
\newcommand{\tbsp}{\rule{0pt}{18pt}}

\title{\bf Monte Carlo Study of the Arrival Time Distribution of
Particles in Extensive Air Showers in the Energy Range 1--100 TeV}

\author{G. Battistoni$^{1}$, A. Ferrari$^{1}$,
M. Carboni$^2$, 
V. Patera$^{3,2}$ \\
{\small $^1$ {\it INFN Milano, 20133 Milano, Italy}} \\
{\small $^2$  {\it INFN Laboratori Nazionali di Frascati, 00044 Frascati (Roma),  Italy}} \\ 
{\small $^3$ {\it   Dipartimento di   Energetica,  Universit\`a ``La Sapienza'', 00185 Roma, Italy}} 
}
\date{}
\begin{document}

\maketitle
\begin{abstract}  
A detailed simulation of vertical showers in atmosphere
produced by primary 
gammas and protons, in the energy range 1--100 TeV,
has been performed by means of the FLUKA Monte Carlo code, 
with the aim of studying the time structure
of the shower front at different detector heights.
It turns out that the time delay distribution can be fitted
using few parameters coincident
with the distribution central moments.
Such parameters exhibit a smooth behaviour as a function of energy.
These results can be used both for detector design  and 
for the interpretation of the existing measurements.
Differences in the time structure between gamma and proton induced showers
are found and explained in terms of the non--relativistic component
of extensive air showers.
\end{abstract} 

\vspace{2.0cm}
\begin{center}
{\it Submitted to ``Astroparticle Physics''}
\end{center}

\newpage

\section{Introduction}
\label{sec:intro}
Recent papers concerning calculations of Extensive Air 
Showers (EAS)\cite{ben} 
include results on time delay of particles from
the shower front, in
view of the comparison with existing recent
measurements\cite{cover,grexpar,eastop}.
As a matter of fact, the first measurements of the structure
of the EAS front were attempted by Bassi, Clark and Rossi in
1953\cite{bassi}, but many other have contributed\cite{linsl1}-\cite{Khris}.
As far as the many simulation works on EAS are concerned,  a part from
the quoted ref. \cite{ben}, only a limited
fraction of them has considered the question of the time
structure\cite{Grieder}-\cite{Rebel}. 
\par
The interest in this topic has been renewed by recent
experimental data concerning the detection of
an anomalous delayed component\cite{ambrosio}.
In this framework it is therefore important to achieve a better and more
detailed knowledge of  this time structure. 
Furthermore, eventual fast simulation tools based of simple parametrisations constructed on this 
knowledge can be helpful not only for analysis, but also for design 
of new detectors, trigger evaluation, et cetera. 

When a full shower simulation is considered,
the structure of the time front depends to a large
extent on the details of particle transport algorithms.
Therefore, it can be useful to consider the outputs 
of very detailed codes, and compare it with the results of the codes
optimized for the cosmic ray physics, which often
contain simplifications in order to reduce the computer time.
A valuable study of the time structure of Extensive Air Showers, for  very
high energies, based on the use of the CORSIKA code\cite{corsika} has
been presented in \cite{grexpar}.
\par
Here we present the study of the time structure of the e.m.
secondary component generated by photon or proton, as  calculated 
mainly with the FLUKA96 code. We have already used such code in
\cite{ciro}, where we published the parametrisations of
e.m. size resulting from photon and electron induced sub-showers.
Such code has already been successfully used in other cosmic ray 
applications\cite{roesler,taup97}.
\par
We have explored primary energy from 100 GeV up to 100 TeV, 
and injected at the top of the atmosphere. 
This energy range is of particular interest for gamma-astronomy purposes, 
mainly in the chance of 
possible differences in the time structure between gamma and proton
initiated showers.

After describing the MC set-up, we discuss the search for the
best functional shape to fit the time distribution, and then results
are discussed. Numerical results of characteristic parameters are given
for different energies, secondary particles and detector
altitudes.

\section{The MC codes}
\label{sec:mc}
Among the shower programs available 
in the High Energy Physics community, we have
mainly considered FLUKA96\cite{fluka}, although in some cases we have
also used GEANT version 3.21 \cite{geant}, 
using the interface to the FLUKA
package (1992 version) for hadronic interactions.
It must be stressed that GEANT-FLUKA and FLUKA96 are not the same thing, 
for different reasons:
the e.m. simulation code is different\cite{emf},
particle transport algorithms are different and, most of all,
the FLUKA interface of GEANT contains only part 
(updated to 1993) of the hadronic interaction model of full FLUKA96.

In particular, the full FLUKA96 code employs transport algorithms
with refined path length corrections\cite{mult} associated to 
multiple scattering, which are essential in problems involving low energy 
electrons and positrons.
Furthermore, FLUKA96 is completely in double precision, 
thus allowing a much more accurate  definition of a finely  
segmented geometry set--up which extends itself up to many tens 
of kilometres. Another advantage coming from double precision is 
the timing accuracy  even below 1 ns. 
An additional advantage of the full FLUKA96 package is the 
inclusion of very detailed models of nuclear excitation and low energy
neutron transport. As we
shall see in the following, this last topic has some importance 
for the argument of
this paper.
The two different codes adopted in this work provide slightly different
results, in particular for log$_{10}{t/ns}\le$ 0, because of the
insufficient precision of the GEANT transport algorithms. However, for 
log$_{10}{t/ns}$ greater than that, the differences in the bulk of arrival
time distributions are substantially within statistical fluctuations.
For all these reasons, in the following sections we shall limit ourselves
to the discussion of the results obtained with FLUKA96.

The atmosphere is defined by a stack of box volumes
of rectangular basis and thickness increasing with the height above the
sea level. Any volume corresponds to a depth of 
$\sim$ 24 g/cm$^2$. In each box the density is uniform, and it is
chosen in such a way that an approximation to
the standard U.S. atmosphere is performed according to the
Shibata fit\cite{atmosphere}.
The chosen depth granularity in our approximation 
is about one half of radiation length (37.66 g/cm$^2$) 
in air.
We have limited the top of atmosphere 
at 1 g/cm$^2$,
and the bottom is at 1025 g/cm$^2$.
Particles have been injected at the altitude H = 45.445 km above the sea level.
We have used the same elemental composition at all depths.
The kinetic energy cut for secondary charged particles has been fixed 
to to 1 MeV, while for secondary photons we have chosen 0.5 MeV in order to
include the contribution from $e^+e^-$ annihilation.
Particles have been recorded at three different detector heights,
corresponding respectively at sea level, 1000 and 2000 m above sea level.
These altitude values have been chosen in order to apply our work mainly
to the case of the Gran Sasso laboratory site.
For our investigation, we generated only vertical showers,  
at different log--spaced energies: 
100, 177, 316, 562, 1000, 1778, 3128, 5623, 10000, 17780, 
31620, 56230, 100000 GeV.

\section{Analysis} 
\label{sec:ana}
 
We present results for different groups of secondary particles.
We have considered separately photons, electrons ( both $e^+$ and $e^-$ )
and other charged particles (muon, pions, etc.). The last group of secondary 
is relevant only in the case of
primary hadrons. It must be noticed in fact that 
the adopted codes did not activate the
hadro-production by photons.
\par
As already done by other authors\cite{eastop} we defined 
as relevant variable, the time delay $t$
with respect to a spherical front moving with light speed $c$, 
originating from a fixed injection point the atmosphere, 
{\it i.e.} we are giving the shower disk deviation from such a spherical front.
The exact definition of our delay variable $t$ is given by:
\begin{equation}
t = t_{arrival} - \frac{\sqrt{H^2+r^2}}{c}
\end{equation}
where $r$ is the distance from the shower core, and $t_{arrival}$ is the
arrival time of the considered secondary particle, as
calculated by the code, and H is the injection point height in the atmosphere.
The injection of the primary particle occurs by definition at zero time.
\par
We have recorded the delay
distribution in different radial bins from the impact point of
the shower axis for each different group of secondary particles. 
On the basis of the features of the lateral distribution of secondary
particles, we use a logarithmic spaced radial zones, for a total number 
of 9 regions:
\begin{itemize}
\item{I}: 0$\div$4.64 m
\item{II}: 4.64$\div$10.00 m
\item{III}: 10.00$\div$21.54 m
\item{IV}: 21.54$\div$46.42 m
\item{V}: 46.42$\div$100.00 m
\item{VI}: 100.00$\div$215.44 m
\item{VII}: 215.44$\div$464.16 m
\item{VIII}: 464.16$\div$1000.00 m
\item{IX}: 1000.00$\div$2154.42 m
\end{itemize}

The contribution at distance exceeding the limit of region 9 are negligible
for the considered energies.

\subsection{The arrival time probability distribution function}

We paid a particular attention to the choice of a suitable
function which can be used to fit the delay distribution
in any of the considered radial regions.
If there is no interest in time delays larger than tens of nanoseconds,
usually a good choice
in experimental analysis is the gamma function, 
as in ref.\cite{ben} or \cite{cover}:

\begin{equation}
\Gamma(t) = A\cdot t^{\beta} e^{-t\cdot \alpha}
\end{equation}

However, the authors often note how
the gamma function cannot account for the long tail of the experimental
delay distribution\cite{cover}.

Our simulations also exhibit such long tails, but we have found that 
the use of a log-normal function allows a fairly satisfactory reproduction
of the simulated data (at least when no smearing effects 
due to finite resolution of detector are included)
 in any of the aforementioned
radial regions, up to a very long delay time ($\simeq 10 \mu$s):
\par
\begin{equation}
f(t) =\frac{A}{t} \cdot e^{\left[ -\left( \log{t}-B\right)^2 /C \right]}
\end{equation}

The $B$ and $C$ parameters of such a distribution 
(essentially related to the mean delay
and to its r.m.s) are also found to follow simple evolution as a function
of shower energy and radial distance. 
\par
The preference for the log--normal behaviour could reflect some intrinsic
feature of the underlying processes. According to ref.\cite{eadie},
the log--normal distribution arises whenever we are in presence of 
a variable whose value takes a random proportion of that of the previous
step in the stochastic process. However, we must stress that our
choice of approximating function was an heuristic fact, 
aiming to the best approximation of the actual distribution. 
In fact we did not adopt any mathematical model as input for the shower 
development, thus 
we merely state that the log-normal distribution provides a better
numerical approximation than the gamma function to the phenomenological delay 
distribution.
Therefore, also the concept of goodness of
fit has to be somewhat relaxed in the present discussion with respect
to a rigorous statistical context.
\par
\begin{figure}
\begin{center}
\mbox{\epsfig{file = 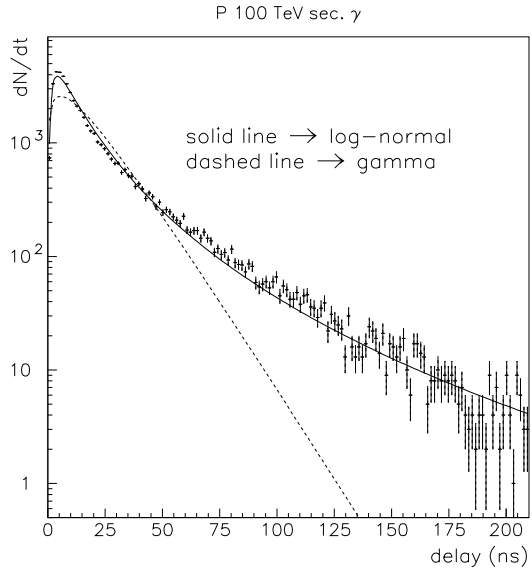,width=8.cm } } 
\caption{Log--normal  and Gamma 
fit to the delay distribution of 
secondary $\gamma$'s and from primary
100 TeV protons, 
as detected at 2000 m a.s.l. in the V radial region
(stand-alone FLUKA96 simulation.\label{fi:1}}
\end{center}
\end{figure}

As an example, in Fig. \ref{fi:1} we show the time delay distribution 
for secondary gamma recorded at 2000 m a.s.l. in the radial 
region V, as produced by primary protons of energy 100 TeV, fitted 
up to $200 ns$ of delay, with log-normal and gamma functions.
\par

The differences are self evident and can be noticed at different
energies, radial regions, etc.
\par
\begin{figure}
\begin{center}
\mbox{\epsfig{file = 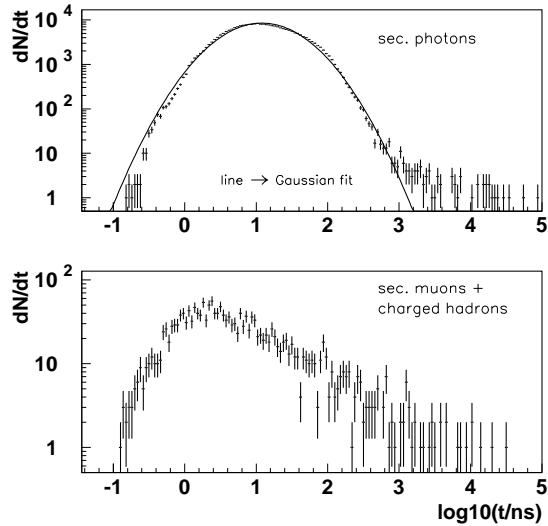,width=8.cm } } 
\caption{
Above: Gaussian 
fit to the $\log_{10}{t}$ distribution of 
secondary $\gamma$'s and from primary
100 TeV protons, 
as detected at 2000 m a.s.l. in the V radial region
(stand-alone FLUKA96 simulation).
Below: $\log_{10}{t}$ distribution of secondary muons
and charged hadrons, for the same primary energy and
distance from the shower core.
\label{fi:1ter}}
\end{center}
\end{figure}

After this conclusion, we find more convenient to re--express all the
results plotting $\log{t}$ instead of $t$ and then fitting to a gaussian,
as shown in the upper part of Fig. \ref{fi:1ter}, which refers to the
same case of Fig. \ref{fi:1}.
The distribution for secondary $e^+e^-$ has practically log-normal shape.
It is also interesting to look at the distribution, for the same energy, radial
bin, etc., when secondary muons and charged hadrons are selected.
This is shown in the bottom part of Fig. \ref{fi:1ter}. 

The resulting distribution, for this primary energy and distance
from the core, is even more asymmetric than that of electrons
or photons. The consequence is that neither the pure gamma or the
pure log--normal distribution succeed in fitting the delay distribution
up to large $t$ values. At smaller distances from the core, also for muons
and charged hadrons, the log--normal approximation works better. 
However, in the following, we shall limit ourselves to the
discussion of the results for $e^+e^-$ and $\gamma$'s, since
their density largely dominates that of muons, or residual charged hadrons.
We recognize that the case of muons and other charged particle would 
deserve a dedicated study, also in view of the fact that it has been
recently
advocated the use of arrival time of muons for measurements related to the
mass composition of primary cosmic rays, using the so called
``Time Track Complementarity'' (TTC)\cite{ttc}.
At present we are not able to perform with our tools a detailed study of
this, since our Monte Carlo code does not allow, in this version, the
treatment of nuclear projectiles and the primary energies required for 
this purposes.

In any case, it already appears that also the log-normal fit
is not sufficient to weight carefully the whole $t$ range.
Such a fit usually takes good care of the bulk of the events, but the
extreme tails of the $\log{t}$ distributions exhibit deviations 
from a perfect gaussian behaviour. 
As a matter of fact, the left tail is very important, since it
corresponds to the early front of the shower, that affects 
the experimental trigger time. The
right tail refers to the very delayed component of EAS.
In order to look for a more precise description of the simulation results, 
we characterise the  $\log_{10}{t}$ distribution and its deviations from
a pure log--normal one by means of a set of parameters.
We find convenient the use of central moments
 ({\it i.e.} the moments calculated around the
mean).
Starting from the values of a finite set of these  moments, the 
distribution can be reproduced ( and directly sampled in Monte Carlo)
 with a sufficient degree of approximation, as demonstrated
in ref. \cite{kendall}.
We consider a variable $\xi$ defined by
$\xi = \frac{\log_{10}(t)-\mu_1}{\sigma}$ where $\mu_1$ and $\sigma$ are the 
average and the R.M.S. of the $\log_{10}(t)$ distribution.
If the $f(\xi)$ distribution function (p.d.f. in the following) can be 
approximated by a standard normal $g(\xi)$ then $f(\xi)$ can be 
expanded in series of the derivatives of the standard normal function:
\begin{equation}
f(\xi) = g(\xi) \{ 
1  +\frac{ \mu_3}{6\sigma^3}
H_3 +\frac{1}{24\sigma^4} \left( \mu_4 - 3\right) H_4 +
\frac{\mu_3^2}{72\sigma^6} H_6 + ...
\}
\end{equation}

Where $H_n$ are the Hermite polynomials of order $n$ and $\mu_n$ are 
the central moments of $f(\xi)$:
\begin{equation}
\mu_n = \sum_{i=1}^N \frac{\left( \log_{10}\left
      ( t_i\right)-\mu_1\right)^n}{N} 
\end{equation}

This is also known as Graham-Charlier expansion (GC in the following).
We have chosen to give the first four moments of $\log_{10}(t)$
distributions. This set seems sufficient to express the features of the 
delay distributions in most cases, but there are some limitations,
as discussed in the literature, which would demand for higher order moments.
For instance when $\mu_4$ increases above a certain limit, 
moments of an order greater than $n$=4 must
be considered in the expansion, otherwise the distribution becomes neither
unimodal neither positive defined in the whole range.
\par
However we must stress how the moments of high order are subject
to statistical fluctuations in case of a small number of entries in a 
distribution. In that case the errors induced by such fluctuations might be
even larger than that due to the omission of such moments.
\par
\begin{figure}
\begin{center}
\mbox{\epsfig{file = 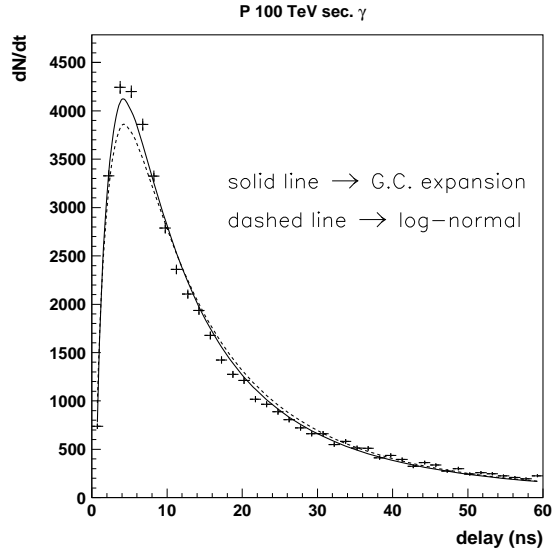,width=8.cm } } 
\caption{Log--normal  and GC
fits to the small delay region of the
distribution of 
secondary $\gamma$'s and from primary
100 TeV protons, 
as detected at 2000 m a.s.l. in the V radial region
(stand-alone FLUKA96 simulation.\label{fi:1bis}}
\end{center}
\end{figure}

To show how the G.C. expansion works, in Fig. \ref{fi:1bis} 
we show the comparison between the
pure log--normal and the GC expansion fits to the short delay region of the
same distribution of Fig. \ref{fi:1} (please note the linear scale!). 
Such an improvement can be seen also at large delays. 
\par
In summary a better $\chi^2$ is obtained but usually the first
two moments resulting from the fit, common to the two functions, are the 
same within the fit errors for the two case.
We stress that, for most of the cases, these first two moments ({\it i.e.} the
  log--normal approximation) are sufficient
to describe the time delay distribution from the experimental point of view
({\it i.e.} $\log(t/ns)>$0), with the noticeable exception of long tails at 
large radial bins for proton showers, as will be discussed in Section 5.

An important outcome of the series expansion in terms of moments is that
it is possible to derive an expression which allows to have a direct Monte
Carlo sample of the
the desired p.d.f. with good accuracy starting from a random number 
normally distributed $\xi$  without any rejection.
This is described and demonstrated in \cite{kendall}.
The relation between variable x to be generated with average $\mu$, standard
deviation $\sigma$ and central moments $\mu_n$, and the 
 normal distributed number $\xi$ is given by:

\begin{equation}
 x = \mu + \sigma\left[\xi + \frac{\mu_3}{6\sigma^3}\left(\xi^2-1 \right) + 
     \frac{\mu_4-3\mu_2^2}{24\sigma^4}\left(\xi^3-3\xi\right) - 
     \frac{\mu_3^2}{36\sigma^6}\left( 2\xi^3-5\xi \right)\right]
\label{xi}
\end{equation}
with the $\mu_n$ giving the n-th central moments of 
the x distribution function. Of course $\mu_2$ = $\sigma^2$.
\par
\begin{figure}[htb]
\begin{center}
\mbox{\psfig{file=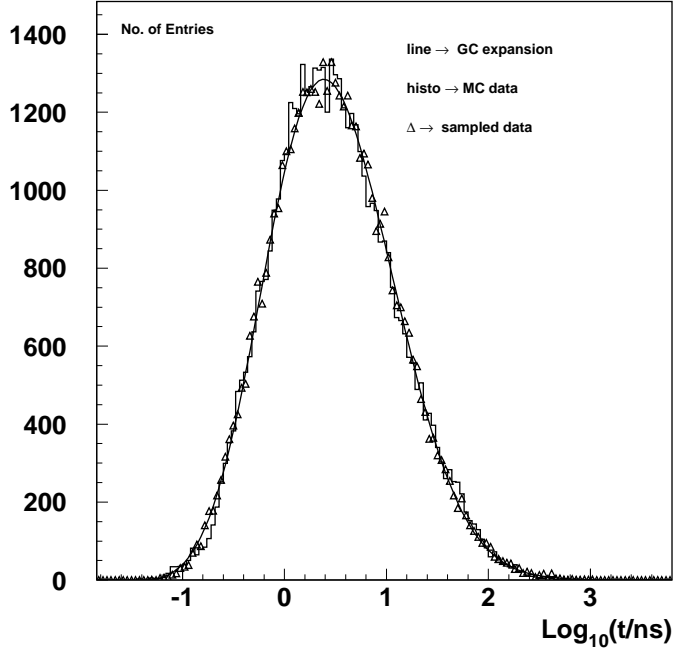,width=10 cm}}
\caption{Comparison between the M.C. data the momentum expansion and the 
sampling described in \cite{kendall}
\label{fig:sampling}}
\end{center}
\end{figure}

An example of the functionality of this sampling method is shown in Fig. 
\ref{fig:sampling} where the M.C. output, the momentum expansion and 
the sampled data from this expansion are superimposed.

\par
Further differences, such as those 
discussed in Section \ref{sec:delay}
can be considered as an estimate of the systematic error 
related to the choice of different MC algorithms and interaction models.

\section{Results}
\label{sec:results}

In this section we present
the M.C. results giving information about the four moments that appear in the
expansion of eq. \ref{xi}, for different
set of primary particle, secondary produced and distance 
from the shower core (radial bin). For practical convenience, instead of
$\mu_2,~\mu_3,~\mu_4$,
in the following we shall use r.m.s 
($\sqrt{\mu_2}$) , skewness ($\mu_3/{\mu_2}^{3/2}$)
and kurtosis ($\mu_4/\mu_2^2 - 3$) in place of $\mu_3$ and $\mu_4$.
For each set of the three quoted parameters, we obtained 
these quantities in two ways: by direct computation of the moments
from the logarithm of the arrival times, and by fitting the $\log(t)$ 
distribution with the GC expansion with the moments used as fit parameters.
A quantitative comparison of this two set of moments is made using the
computed one as fixed parameters of the GC function and fitting the 
$\log(t)$ histogram with only the normalisation as a free parameter.
In most cases, the two sets of resulting moments are in good agreement,
 within the errors, and
also the $\chi^2$ values from the two fits are similar. 
Important exceptions are
those relative to the regions where we had
low statistics. There, the moments obtained from the general fit 
generally give a 
worse approximation of the data than those directly computed.
\par
We will show how the behaviour of the parameters is rather smooth, 
so that all intermediate
cases in radius and detection height can be easily obtained by interpolation.
We shall express the results in two ways: we will consider the moments of
the average time distribution of single secondary particles ({\it i.e.}
the distributions obtained summing the arrival times for each particles
and for all the primary showers), but we shall also give quantitative
information on the statistical fluctuations of these moments from event to
event. Unfortunately, our simulation runs at the highest energies have only
a small number of events, so that while the single particle distribution is
always measured with high accuracy, the fluctuations on an event by event
basis are studied with less precision.

\begin{figure}
\begin{center}
\mbox{\epsfig{file =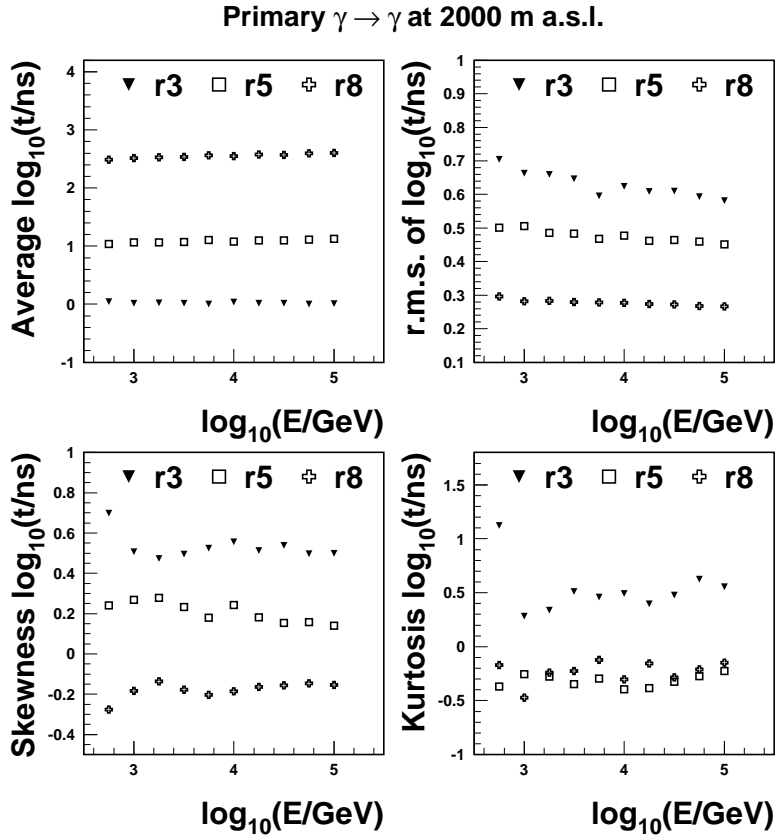,width=12cm } }
\caption{
First central momenta of the arrival time p.d.f. versus $\log(E)$ for 
secondary $\gamma$ from $\gamma$ primary in 3 different radial bins.
Here r3, r5, and r8 stand for radial bin III, V and VIII.
\label{fi:g_g_mom}}
\end{center}
\end{figure}

\subsection{The p.d.f. moments}
\par
The obtained moments\footnote{for the sake of simplicity, we shall use in 
the following the term moment, although r.m.s, skewness and kurtosis are
not 
the true moments} for the distributions recorded
at the observation level of 2000 m. a.s.l. are given in the tables reported
in the Appendix.
As previously stated, we limit ourselves to the results for
secondary $e^+e^-$ and $\gamma$, which dominate the shower size.
We refer mainly to the level of 2000 m a.s.l. since, in 
applications like gamma-astronomy, it is the most favourable among the
three a--priori foreseen observation levels, although higher altitudes would
be even more appealing for this purpose.
The tables of parameters at the other lower
observation levels can be obtained from the authors.

In this section we present a subsample of these results in a graphical
way in order to discuss the essential features. We plot the 
moment values without the corresponding error as derived from the fit.
This is because those errors are dominated by the generated  statistics 
and are lower than the actual shower to shower 
fluctuations, which instead are specifically reported in this paper.
There is also another important comment about these
results. Since the moments have been extracted from a fit, there exists some
degree of correlation among them. This makes even less significant the 
error on a single parameter. We cannot report here the covariance
matrix for all the relevant cases, and the irregularities visible in these
figures are not simply due to statistical fluctuations.
Correlations are such that the evolution of the resulting distribution is much 
smoother than that of single parameters.
   
\begin{figure}
\begin{center}
\mbox{\epsfig{file =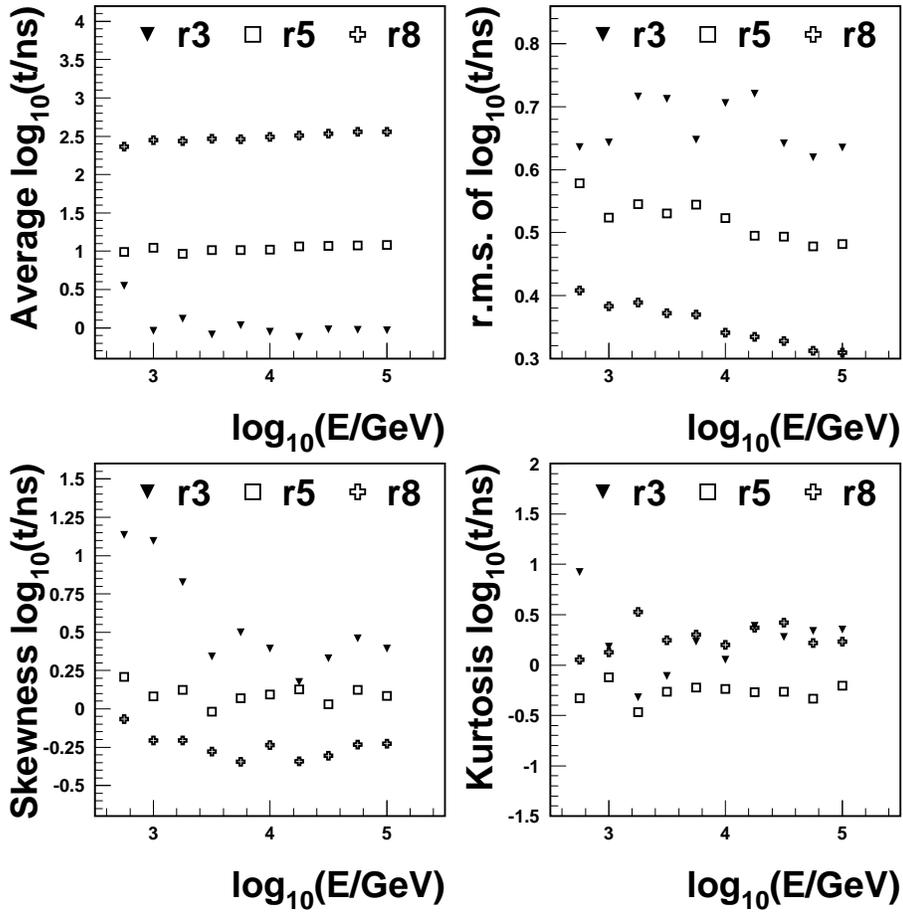,width=14cm } }
\caption{
First central momenta of the arrival time p.d.f. versus $\log(E)$ for 
secondary $\gamma$ from proton primary in 3 different radial bins.
Here r3, r5, and r8 stand for radial bin III, V and VIII.
\label{fi:p_g_mom}}
\end{center}
\end{figure}

 In Fig.  \ref{fi:g_g_mom} the moment values are shown
versus the logarithm of the primary energy for three radial bins,
for the case of showers induced by $\gamma$ primaries and
in Fig. \ref{fi:p_g_mom} are plotted the same 
quantities for proton induced showers. 

The most striking feature is that the
average value of $\log(t)$ is almost independent from primary energy (at
least in the range considered by our simulation), kind of 
primary or secondary, and changes only with the radial bin.
\par
If we adopt the log--gaussian approximation, we remind that, for
this distribution, the correspondence
between $<t>$, $\sigma_t$  and $<\log_{10}\left(t\right)>$, $\mu_2$ 
is given by:
\begin{equation}
\log_{10}<t> = <\log_{10}\left(t\right)> + 
             \frac{\mu_2}{2} \log_e\left(10\right)
\end{equation}
\begin{equation}
 \sigma_t = <t> \sqrt{\left( 10^{\mu_2 \log_e \left(10\right)} - 1\right)}
\end{equation}
The increase in the average delay (which we remind is measured with respect
to a light ray) as a function of distance from the shower core reflects
the fact that moving to the outer regions of the shower, the average
energy of secondary particles decreases, and therefore the deviations
from a straight line trajectory become more important.

The values of $\sqrt{\mu_2}$ are slightly different according to the different 
primary or secondary species, but is substantially constant with respect to the
primary energy for $\gamma$ induced shower and shows a weak dependence from 
this parameter for proton primary. 
We also notice an apparent reduction of the variance (i.e. a shrinkage of
the distribution) as function of the radial distance from the core.
One should not be induced in error, since we plot the variance of
$\log{t}$: it can be verified that transforming back to the $t$
variable, the effective width of the delay distribution (directly related to
the time thickness of the shower disk) increases as a function of distance.
Let us take a numerical example, for a primary gamma of 10 TeV, 
we have a r.m.s. around 0.6
in the third radial bin, when the average is 0.004. 
It means that the 68\%
fraction in $\log{t}$ is in the range 0.18 $\div$ 6.0 ns. Instead, in the
eighth radial bin, the r.m.s. drops  0.28, while the average 
increases to 2.6. Therefore, the 68\% 
fraction in $\log{t}$ is now in the range 118 $\div$ 1350 ns.

\begin{figure}
\begin{center}
\mbox{\epsfig{file =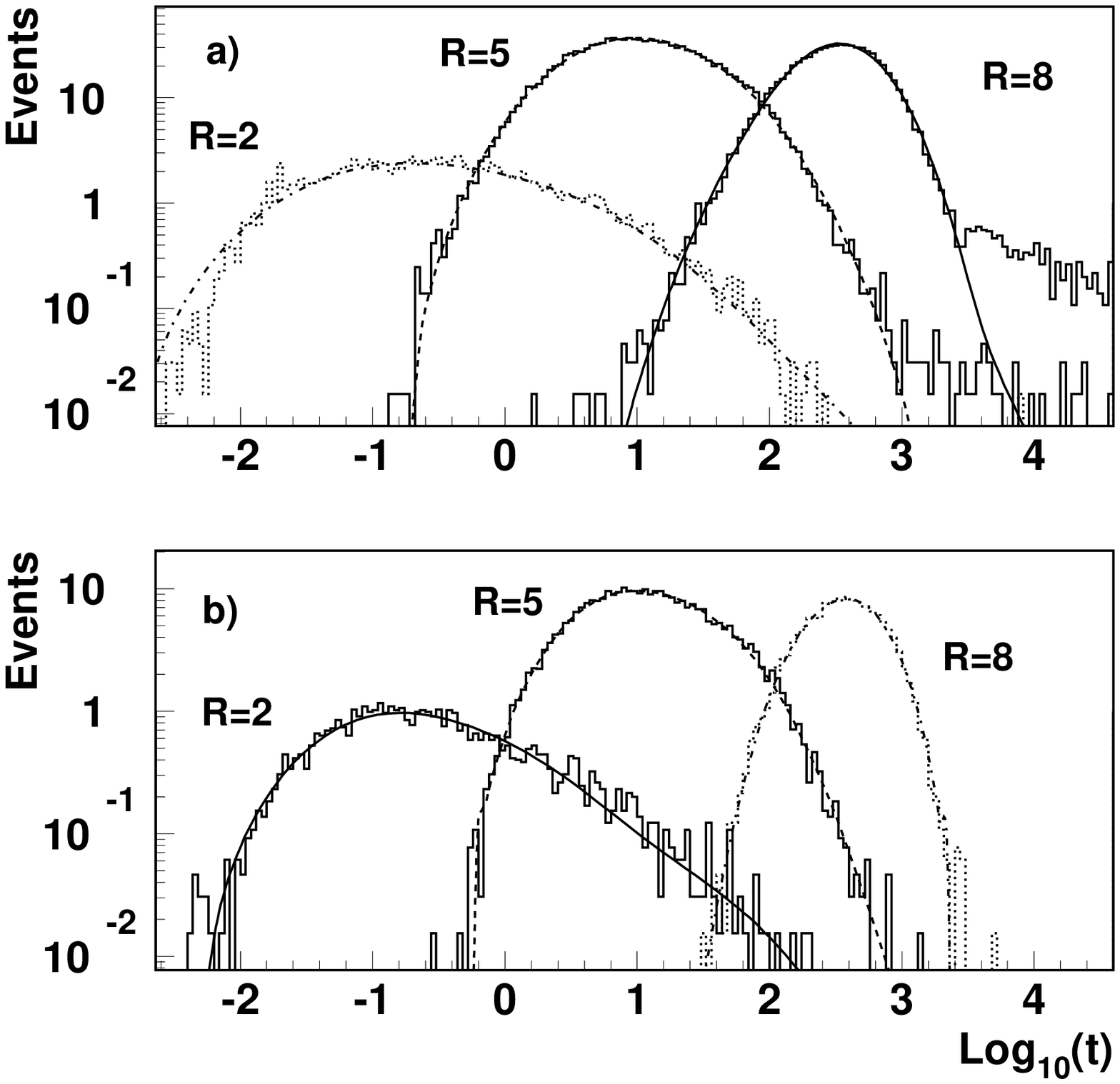,width=14cm } }
\caption{$Log_{10}t$ distributions
at 10 TeV for proton (above) and $\gamma$ (below) showers in few different 
radial bins.
Here r2, r5, and r8 stand for radial bin III, V and VIII.
\label{fi:timevsr}}
\end{center}
\end{figure}

The higher moments are very different for the
two primaries: roughly speaking the proton induced showers have higher third 
and fourth moments, that is the $\log(t)$ 
p.d.f. has a less log-normal shape. 
In order to understand that, but also to give a clearer idea of
the evolution of time delay distributions, we show in Fig. \ref{fi:timevsr}
the $\log{t}$ distributions
at 10 TeV for $\gamma$ and proton showers in few different radial bins.
We can clearly see that while there is a nearly log--gaussian bulk almost
identical in the two cases, an additional tail at high delays appears for 
proton showers, particularly visible at large distance from the shower core.
Our fit is unable to follow completely this tail, which presents itself
as a distinct family of particles. Indeed the population of
these highly delayed particles is only a very small fraction of the 
total number of
arriving e.m. particles. 
We postpone the specific discussion on this phenomenology in 
section \ref{sec:delay}.

\begin{figure}
\begin{center}
\mbox{\epsfig{file =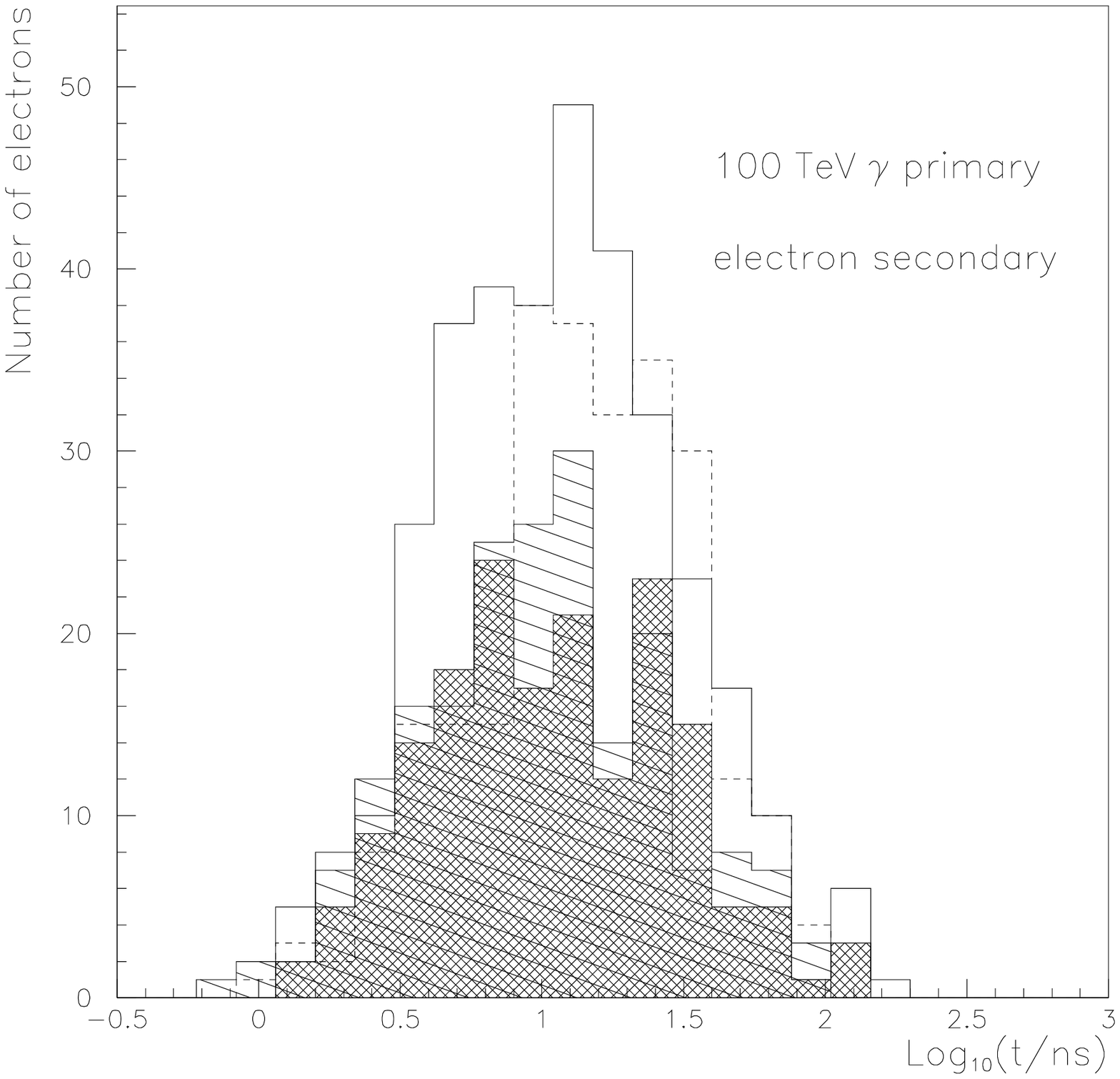,width=14cm } }
\caption{Electrons arrival time distributions for single showers from gamma 
100 TeV primary.
\label{fi:single}}
\end{center}
\end{figure}

\begin{figure}
\begin{center}
\mbox{\epsfig{file =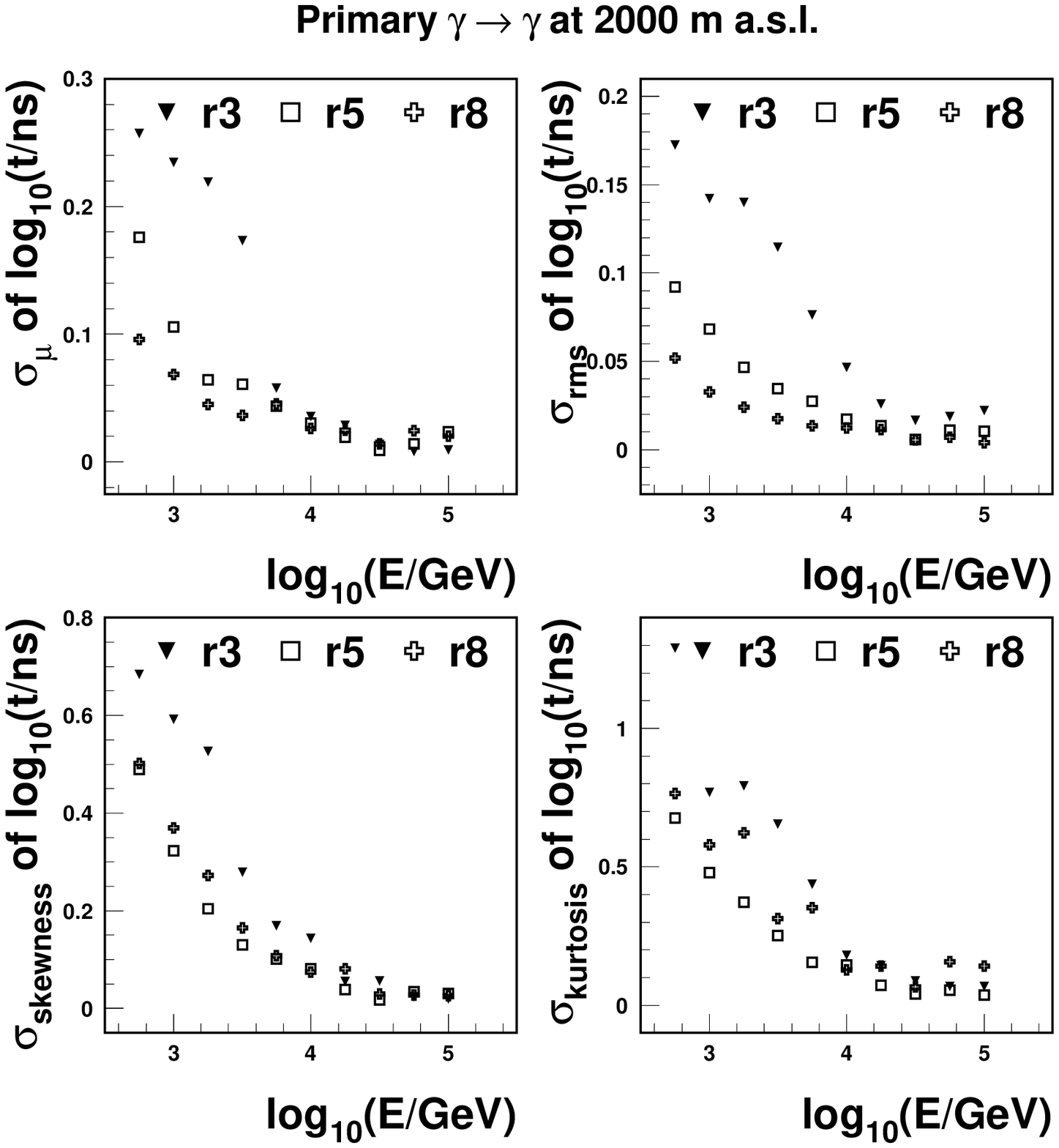,width=14cm } }
\caption{
Fluctuations of first central momenta of the arrival time p.d.f. 
versus $\log(E)$ for secondary $\gamma$ from gamma primary.
Here r3, r5, and r8 stand for radial bin III, V and VIII.
\label{fi:g_g_mfl}}
\end{center}
\end{figure}

\begin{figure}
\begin{center}
\mbox{\epsfig{file =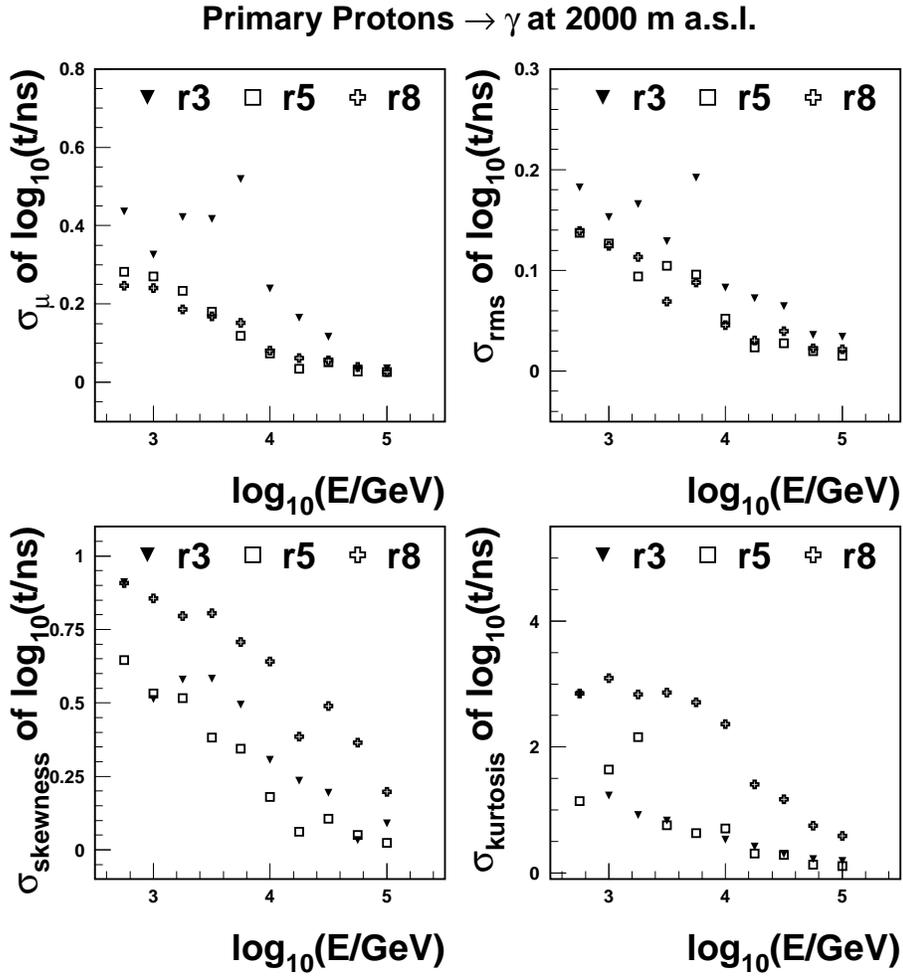,width=14cm } }
\caption{
Fluctuations of first central momenta of the arrival time p.d.f. 
versus $\log(E)$ for secondary $\gamma$ from proton primaries.
Here r3, r5, and r8 stand for radial bin III, V and VIII.
\label{fi:p_g_mfl}}
\end{center}
\end{figure}

Coming back to the general features of the delay 
distribution, we stress that the extracted moments describe 
the average arrival time distribution functions. Obviously, 
remarkable fluctuations can be detected in a event-by-event 
analysis of the generated data. For example, in Fig. \ref{fi:single} we 
superimpose
the electron arrival time distribution for 4 different showers induced by a 
100 Tev $\gamma$ primary. The differences in shape and normalisation for the
different showers are evident. This effect is amplified in the showers induced
by proton primaries. This is highly correlated 
to the large fluctuations 
in the
height of the first inelastic interaction.
In Fig. \ref{fi:g_g_mfl} are
presented the momenta fluctuation for $\gamma$ primaries, to be compared 
with much bigger fluctuation presented in Fig. \ref{fi:p_g_mfl},
corresponding to the proton primaries. As expected,
all the fluctuations shows a decreasing behaviour with the energy.

\section{Discussion of the origin of delayed particles}
\label{sec:delay}

The time structure of $e^+e^-$ and $\gamma$ secondaries
from hadronic and pure e.m. showers shown in Fig. \ref{fi:timevsr} 
is rather similar, as far as the 
nearly log--gaussian bulk is concerned. The main difference consists
in the appearance, in the hadronic showers, of a slow e.m. component
which becomes more and more visible at large distance from the shower core.
Such a component seems to constitute a different family,
really well separated from the major population. It can extend itself 
up to tens of microseconds.
We have investigated in some detail the simulation steps, in order to
understand the physical origin of such a delayed component.
We have learned, of course, that the bulk of our time delay distribution
 come from the relativistic part of the shower,
namely from the e.m.showers coming from $\pi_0$ decay. Instead, the delayed
family
in hadronic showers comes from the de-excitation of air nuclei after the 
interactions with neutron of energy below few tens of MeV. These neutrons
are quite slow, and for this reason separate easily from the relativistic
component. 
Of course, this secondary electro--magnetic component is a small fraction
of a much larger population of neutrons which dominates the shower disc
at large distances. Table \ref{tb:boh2} shows the percentage fraction 
of delayed particles, neutrons, $\gamma$ and electrons, above a radial
dependent time threshold, for 100 TeV proton primary at 2000 m a.s.l.. 
The time threshold for a given radial bin has been chosen in such a way to
approximately define the transition region between the log--normal bulk and
the additional tail.

 \begin{table}
 \begin{center}
 \begin{tabular}{|cccccc|}
  \hline
  Rad. bin 
& time thr. 
& $\frac{N_{delayed}}{N_{total}}\times 10^{-4} $ 
& $\frac{N_{\gamma}}{N_{delayed}}$  
& $\frac{N_{neutrons}}{N_{delayed}} $ 
& $\frac{N_{e^+e^-}}{N_{delayed}}$  \\
\hline
 II & $\geq10^2$ ns&   9.3 &  89.5 (\%)  &  3.5(\%)   & 7.0(\%)  \\
 V & $\geq10^3$ ns&   5.4 &  30.0 (\%)  & 59.0 (\%)  & 4.0(\%)  \\
 VIII & $\geq10^4$ ns& 196.8 &  6.5  (\%)& 93.0 (\%)  & 0.5(\%)  \\
  \hline
 \end{tabular}
 \caption{
Percentage of delayed particle and relative abundance of neutrons, 
photons and electrons for different radial bins, in EAS induced by a 100 TeV 
proton primary.
\label{tb:boh2}}
 \end{center}
 \end{table}

The treatment of nuclear effects, and in particular of de-excitation
processes, is one of the most interesting feature offered by the FLUKA code.
The description of the modelization of these processes is given in
ref. \cite{Ferrari96a} and \cite{Ferrari96b}.
We have to notice that also in the FLUKA interface of GEANT 3.21 a nuclear
evaporation model is present, but it much more simplified and approximated
with respect to that of the quoted references. 
We are aware that this new evaporation model is now also interfaced to
CORSIKA\cite{corsika}, version 5.60, when the DPMJET interaction
model\cite{dpmjet}  is chosen. 
However such a package in the frame of CORSIKA is
not yet available for public distribution.

Here we try to summarize the possible different low-energy
reactions which
can contribute to the process under investigation.
\begin{itemize}
\item[a)]  capture reactions $(n,\gamma)$:
very probably these are not the relevant phenomena in our case, since 
these would give photons with much larger delays than those considered
in our distributions. However, in general they have a small cross section.
This reaction can be relevant for Ar (630 mbarn) through the process
$^{40}$Ar$(n,\gamma)^{41}$Ar.
On this respect we have to make the following remark. The present
 simulation has
been performed in the case of a dry atmosphere, but in case of a relevant
presence of humidity in air or near the soil (in presence of snow for
instance), the build up of capture reactions can be relevant due to the
thermalization of neutrons on H nuclei.

\item[b)] reactions $(n,n')$, where the neutron is scattered 
leaving the nucleus in an excited state. 
The residual nucleus will decay to the ground state emitting 
a cascade of photons between the different levels.
This phenomenon is one of the most important sources of delayed photons,
since an energy just above the first excited level is enough.
For N, the first three excited levels are at 2.319, 3.948, 4.915 MeV.
For Ar they are at 1.461, 2.121, 2.524 MeV, while for O they are at
6.049, 6.130, 6.919 MeV 
(therefore is less relevant in the O case).
The excitation of higher levels is usually negligible.
The lifetime of these excited levels is quite short (ns scale or less), so
that
these photons are promptly emitted. Furthermore, the model makes use of
an isotropic angular distribution. This is not very different 
from reality.

\item[c)] There are also reactions with production of other particles together
with additional $\gamma$'s.
We can quote, in order of increasing energy threshold:
\par
$(n,p)$; among these, an important 
exception is the celebrated $^{14}$N$(n,p)^{14}$C, 
which does not produce $\gamma$'s;
\par
$(n, 2n)$. This is practically closed for O nuclei below about 20 MeV, 
but is open
for the other nuclei of air even at lower energies;
\par
$(n,\alpha)$;
\par
$(n,d)$;
\par
$(n,t)$;
\par
$(n,kp)$, with  $k$ produced protons,
\par
$(n,k\alpha)$, with $k$ produced $\alpha$'s
\par
For each of these processes, the cross section rises quite fast above 
the threshold. 

\end{itemize}
The reason why we notice the delayed family at large distance is that
the low energy neutrons can be found more easily in the periphery of the
shower, where energy is degraded. Furthermore, this non relativistic
component
is obviously much more smeared in space than the relativistic one.
Of course, the delayed neutron component can give much larger
signals in a given detector (for instance in a scintillator array)
with respect to the associated secondary
e.m. component. However, we would like to stress that in case of nuclear
projectiles (a case outside the scope of the present simulations) we shall
have also excitation of the nuclear fragments coming from the
projectile. Unlike the case of the excitation of the target nuclei,
the photons (and other particles) following the de-excitations of the
fast moving fragments will be Lorentz--boosted in the laboratory frame, so
that to contribute in a more efficient way to the energy deposition.
In particular we advance the hypothesis that they can give rise to
subshowers having some delay with respect to the first particles
in the EAS disc.
For instance, we suggest that this 
could be the basis for an explanation of the observation of a delayed 
component\cite{ambrosio} in VHE Extensive Air Showers, as
observed in the COVER-PLASTEX detector in the GREX array\cite{cover}.

We wish to point out that the inclusion of
these processes in simulation codes for high energy cosmic ray physics 
has not yet become a common practice.
Therefore, these phenomena might have escaped from other simulation
studies, while the existence of such delayed particles might have some
experimental relevance.

\section{Comparison with experiments}
\label{sec:compexp}

\begin{figure}
\begin{center}
\begin{tabular} {c}
\mbox{\epsfig{file=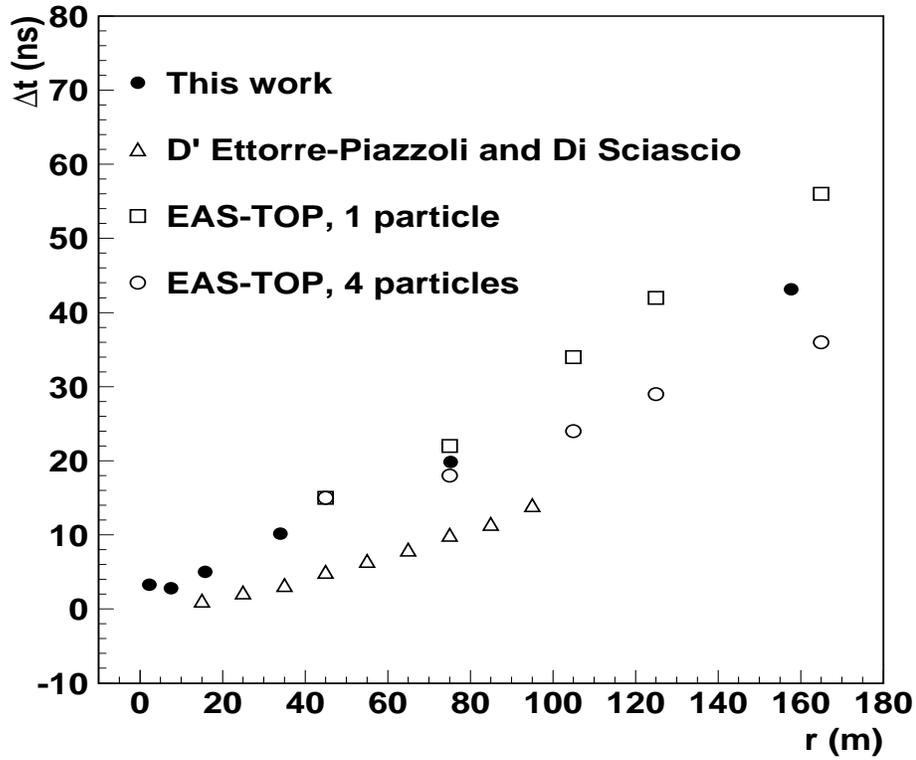,
width=12.cm,height=10cm } } \\
\mbox{\epsfig{file=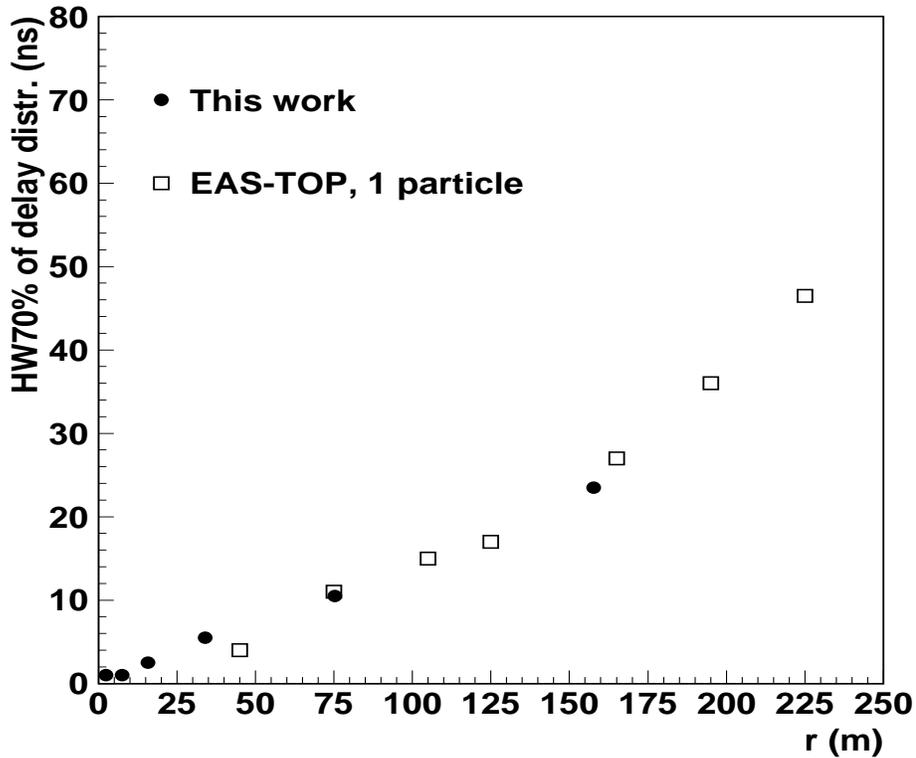,
width=12.cm,height=10cm } }
\end{tabular}
\caption{
Average particle delay from the first arrived particle and width of the 
delay distribution at 70 \% of the peak vs. the 
shower axis distance for $100 TeV$ primary protons (see text for details).
\label{fi:timedel}}
\end{center}
\end{figure}

We compare our results with the EAS-TOP\cite{eastop} array experimental data. 
In Fig. \ref{fi:timedel} we plot the
average secondary delay from the first particle, with respect to the 
radial bin. This values refer to the photons produced by 
a 100 TeV, proton initiated EAS as detected at 2000 m a.s.l. 
The EAS-TOP signal is dominated by the secondary electrons, but  
we remark that, at least in this range of parameters, the differences with
respect to the time distribution of secondary electrons appear to be 
very small. On the other
hand the number of photons is much larger\cite{ciro}, and this allows
a reduced statistical uncertainty.
The average value we obtain seems to reproduce the EAS-TOP experimental
data. These are divided in 2 samples, corresponding to the cases in which
they had at least 1 or 4 particles in each of
the triggered sub-detectors.
The event class corresponding to at least 1 particle in the detector should
correspond to a primary energy close to the trigger threshold of
EAS--TOP ($\simeq$ 100 TeV, mostly from 
proton primary). At distances from the core lower than 100 m, the two
classes are not distinguishable.
In the same figure we also show that
our results are systematically higher than those of ref.\cite{ben}. There
the calculation was done only for $gamma$ primaries, but our results, in
the same range of distance from the shower core, do not exhibit substantial
differences between the two species of primaries.
Also the delay distribution width (that is the shower disk thickness), 
is well reproduced by our  simulation: in Fig. \ref{fi:timedel} is reported 
the  width of the delay distribution as measured at 70\% of the 
height of the peak, for  different radial bins. 
The agreement between Monte Carlo and the
EAS-TOP data is very good.
In spite of the goodness of the simulation results we want to stress 
the effect of a finite detector resolution 
on the simulated arrival time of the secondaries. In particular,
in the first radial bins (near the axis of the shower), the arrival 
time distribution are peaked at time values  which can be lower
than the typical 
experimental time resolution. Thus the result of the folding of the 
simulated p.d.f. with the experimental error distribution drastically 
modifies the distribution in these radial bins, producing a flattening
at arrival times close to the time resolution value. 
\par
\begin{figure}
\begin{center}
\mbox{\epsfig{file =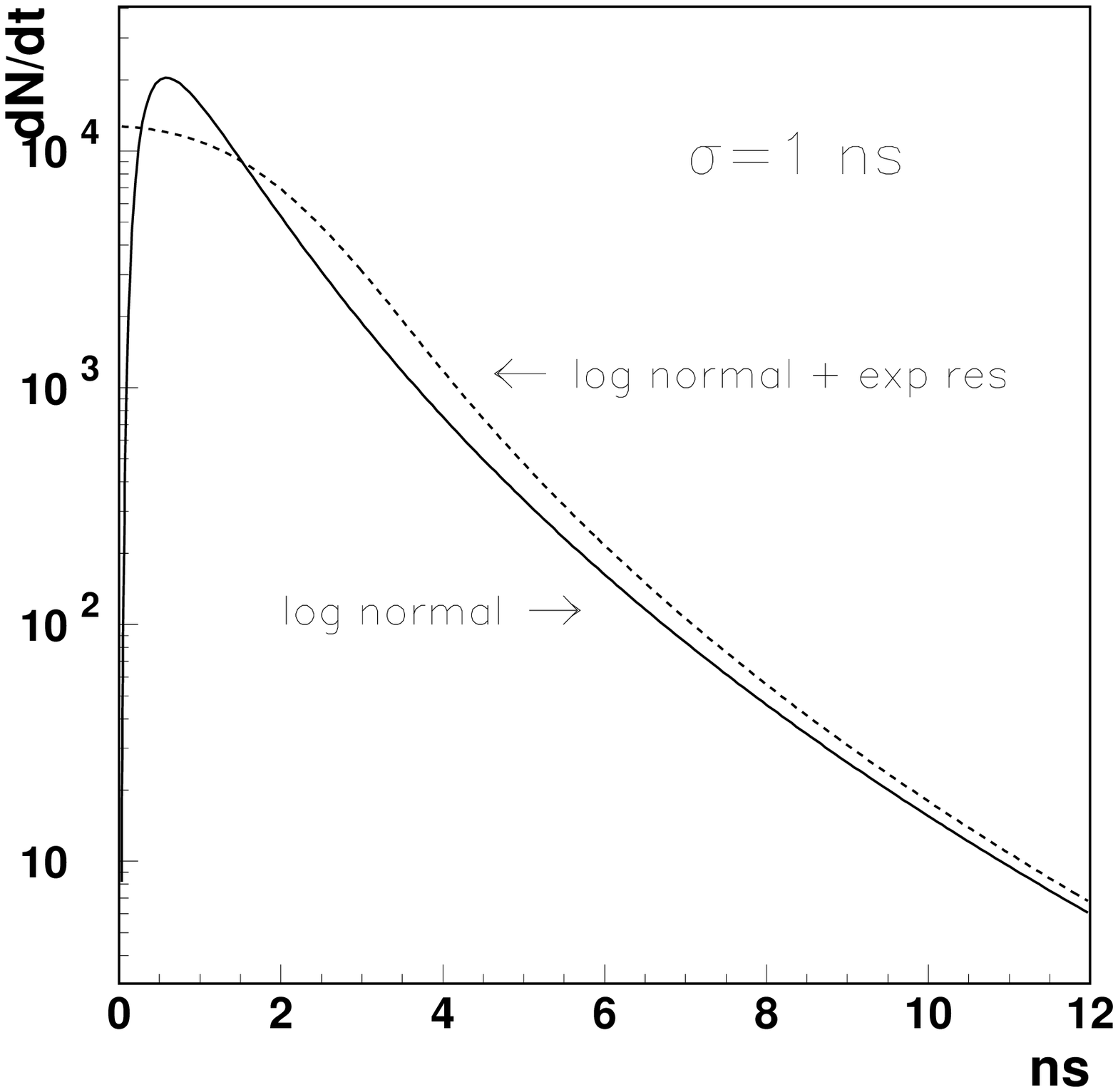,width=14cm } }
\caption{Effect of the folding of a gaussian experimental resolution
  ($\sigma$= 1 ns) with a log--normal distribution, similar to
  the delay distribution expected at 2000 m a.s.l for secondary
photons from primary gamma, in the 2nd radial bin.
\label{fi:confamb}}
\end{center}
\end{figure}

This effect is shown in Fig. \ref{fi:confamb} where we
plotted  a log-normal function with  $<\log_{10}(t/ns)>=-0.54 $ and
$\sigma_{log_{10}(t/ns)}=0.75$ (similar to
  the delay distribution expected at 2000 m a.s.l for secondary
photons from primary gamma, in the 2nd radial bin),
 superimposed to the same 
distribution folded with a gaussian
error function with $\sigma_{exp}=1$ ns. The effect of the experimental 
resolution is clearly seen on the bulk of the events, even if the average
value of the arrival time is bigger then the resolution value. 

We remark also that usually the experimental results are expressed as particle
delay distribution as measured since the arrival of the
experimental trigger. This is different from
the delay time from the light cone that we used in M.C.
 These two  distributions are almost equal 
if the shower is so populated that the first triggering particle has a minimum 
delay from the light cone, but can be quite different for low statistic
showers, where the triggering particle may arrive after a sizeable
delay from the light cone. In this case the M.C. distribution are 
biased toward the longer delays with respect to the 
experimental results by an amount of time given by the average delay
of the first particle.

\section{Conclusion}
\label{sec:concl}
We have simulated the time delay of vertical Extensive Air Showers by means of 
the FLUKA Monte Carlo code.
We have shown how the logarithm of the time delay distribution of a shower
front can be easily expressed as a gaussian expansion in terms of a few
parameters. Such parameters exhibit a smooth behaviour as a function of
energy, thus allowing the extrapolation to higher energies of the presented 
Monte Carlo simulation results.
\par
The proposed expansion allows the construction of fast algorithms useful
in detector design and data interpretation, when
the running of a complete detailed simulation would result in
unnecessary heavy job.
Differences between protons and gamma initiated showers are observed.
In particular, hadronic showers exhibit a delayed component at large
distance from the shower axis. These differences have been interpreted as
 due to 
effects coming from non relativistic particles in the shower. 
In particular, we found that the de--excitation of air nuclei 
following the interaction with low energy neutrons are the 
essential contributions to the highly delayed e.m. component 
in hadronic showers.
We remark how this can be obtained at simulation level, only with 
very specialized and detailed codes, not commonly used so far in 
cosmic ray physics. In particular situations, these results can 
be used as an interpretation basis for the understanding of 
apparently anomalous delays.

\section*{Acknowledgements}
We wish to thank the colleagues of MACRO\cite{macro} experiments at Gran Sasso for
having stimulated this research. In particular we are grateful to
M. Spinetti for his support. We also acknowledge many interesting
discussions with M. Ambrosio, G. Navarra and P. Vallania concerning the 
comparison to experimental data.

\clearpage

\section*{Appendix: Tables of Moments and of their Fluctuations}

\tablefirsthead{
\\
  \multicolumn{5}{c}{Primary $\gamma$, Sec. $\gamma$, h$_{det}$=2000 m a.s.l.} 
\tbsp  
}

\tablehead{\hline 

  \multicolumn{5}{l}{\small\sl continued from previous page}

\\ \hline\tbsp  }

\tabletail{\hline
  \multicolumn{5 }{r}{\small\sl continued on next page}
\\\hline}

\tablelasttail{\\ }

\topcaption{
Central momenta for the distribution function of 
$\log\left(\frac{t}{ns}\right)$ for secondary $\gamma$ produced at 2000 m a.s.l. by
a $\gamma$ primary, as obtained
  from the FLUKA stand-alone simulation}

\begin{center}
\begin{supertabular}{ccccc}
\\ 
 \hline
\multicolumn{5}{c}{ E$_0$ =    562. GeV} \\
\hline
 Rad. bin & average & sigma  &skewness & kurtosis \\
\hline
  1 &  -.77$\pm$   .45 &  1.03$\pm$   .28 &    .3$\pm$    .7 &   3.2$\pm$   1.1 \\
  2 &  -.35$\pm$   .30 &   .83$\pm$   .19 &    .5$\pm$    .6 &   3.0$\pm$    .7 \\
  3 &   .05$\pm$   .26 &   .71$\pm$   .17 &    .7$\pm$    .7 &   4.1$\pm$   1.3 \\
  4 &   .51$\pm$   .25 &   .61$\pm$   .15 &    .3$\pm$    .6 &   3.1$\pm$    .9 \\
  5 &  1.04$\pm$   .18 &   .50$\pm$   .09 &    .2$\pm$    .5 &   2.6$\pm$    .7 \\
  6 &  1.54$\pm$   .12 &   .44$\pm$   .07 &    .0$\pm$    .4 &   2.5$\pm$    .6 \\
  7 &  2.04$\pm$   .10 &   .36$\pm$   .05 &   -.2$\pm$    .4 &   2.6$\pm$    .6 \\
  8 &  2.49$\pm$   .10 &   .30$\pm$   .05 &   -.3$\pm$    .5 &   2.8$\pm$    .8 \\
  9 &  2.91$\pm$   .11 &   .24$\pm$   .07 &   -.2$\pm$    .6 &   3.0$\pm$    .7 \\

\hline
\hline
\multicolumn{5}{c}{E$_0$ =  1000. GeV} \\ 
\hline
 Rad. bin & average & sigma  &skewness & kurtosis \\
\hline
  1 & -1.12$\pm$   .41 &  1.01$\pm$   .23 &    .5$\pm$    .4 &   3.1$\pm$   1.0 \\
  2 &  -.37$\pm$   .30 &   .79$\pm$   .24 &    .3$\pm$    .5 &   3.0$\pm$   1.0 \\
  3 &   .02$\pm$   .23 &   .66$\pm$   .14 &    .5$\pm$    .6 &   3.3$\pm$    .8 \\
  4 &   .53$\pm$   .19 &   .58$\pm$   .09 &    .4$\pm$    .4 &   3.1$\pm$    .7 \\
  5 &  1.06$\pm$   .11 &   .51$\pm$   .07 &    .3$\pm$    .3 &   2.7$\pm$    .5 \\
  6 &  1.56$\pm$   .08 &   .42$\pm$   .05 &    .0$\pm$    .2 &   2.5$\pm$    .4 \\
  7 &  2.06$\pm$   .07 &   .35$\pm$   .03 &   -.2$\pm$    .3 &   2.6$\pm$    .4 \\
  8 &  2.51$\pm$   .07 &   .28$\pm$   .03 &   -.2$\pm$    .4 &   2.5$\pm$    .6 \\
  9 &  2.92$\pm$   .08 &   .23$\pm$   .05 &    .0$\pm$    .5 &   2.6$\pm$    .7 \\

 \hline
 \hline
\multicolumn{5}{c}{ E$_0$ =   1778. GeV} \\
\hline
 Rad. bin & average & sigma  &skewness & kurtosis \\
\hline
  1 & -1.16$\pm$   .32 &   .99$\pm$   .24 &    .2$\pm$    .9 &   2.9$\pm$   1.6 \\
  2 &  -.47$\pm$   .33 &   .79$\pm$   .17 &    .6$\pm$    .5 &   4.3$\pm$   1.1 \\
  3 &   .02$\pm$   .22 &   .66$\pm$   .14 &    .5$\pm$    .5 &   3.3$\pm$    .8 \\
  4 &   .55$\pm$   .12 &   .57$\pm$   .08 &    .3$\pm$    .3 &   3.1$\pm$    .6 \\
  5 &  1.06$\pm$   .06 &   .49$\pm$   .05 &    .3$\pm$    .2 &   2.7$\pm$    .4 \\
  6 &  1.58$\pm$   .05 &   .42$\pm$   .03 &    .0$\pm$    .2 &   2.5$\pm$    .3 \\
  7 &  2.07$\pm$   .05 &   .34$\pm$   .02 &   -.2$\pm$    .2 &   2.5$\pm$    .3 \\
  8 &  2.53$\pm$   .05 &   .28$\pm$   .02 &   -.1$\pm$    .3 &   2.8$\pm$    .6 \\
  9 &  2.94$\pm$   .06 &   .23$\pm$   .05 &    .0$\pm$    .5 &   2.9$\pm$    .7 \\

 \hline
 \hline
\multicolumn{5}{c}{ E$_0$ =   3162. GeV} \\
\hline
 Rad. bin & average & sigma  &skewness & kurtosis \\
\hline
  1 & -1.04$\pm$   .37 &   .94$\pm$   .20 &    .8$\pm$    .6 &   4.5$\pm$    .9 \\
  2 &  -.53$\pm$   .25 &   .73$\pm$   .15 &    .6$\pm$    .4 &   4.1$\pm$    .9 \\
  3 &   .02$\pm$   .17 &   .65$\pm$   .11 &    .5$\pm$    .3 &   3.5$\pm$    .7 \\
  4 &   .55$\pm$   .08 &   .56$\pm$   .05 &    .4$\pm$    .2 &   2.9$\pm$    .4 \\
  5 &  1.07$\pm$   .06 &   .48$\pm$   .03 &    .2$\pm$    .1 &   2.7$\pm$    .3 \\
  6 &  1.59$\pm$   .04 &   .41$\pm$   .02 &    .0$\pm$    .1 &   2.4$\pm$    .2 \\
  7 &  2.08$\pm$   .04 &   .34$\pm$   .02 &   -.2$\pm$    .1 &   2.6$\pm$    .2 \\
  8 &  2.54$\pm$   .04 &   .28$\pm$   .02 &   -.2$\pm$    .2 &   2.8$\pm$    .3 \\
  9 &  2.96$\pm$   .05 &   .23$\pm$   .03 &   -.1$\pm$    .3 &   3.0$\pm$    .5 \\

 \hline
 \hline
\multicolumn{5}{c}{ E$_0$ =   5623. GeV} \\
\hline
 Rad. bin & average & sigma  &skewness & kurtosis \\
\hline
  1 & -1.44$\pm$   .32 &   .95$\pm$   .17 &    .1$\pm$    .5 &   4.2$\pm$    .7 \\
  2 &  -.54$\pm$   .13 &   .66$\pm$   .09 &    .7$\pm$    .4 &   4.0$\pm$    .8 \\
  3 &   .00$\pm$   .06 &   .60$\pm$   .08 &    .5$\pm$    .2 &   3.5$\pm$    .4 \\
  4 &   .57$\pm$   .04 &   .53$\pm$   .04 &    .4$\pm$    .1 &   3.1$\pm$    .2 \\
  5 &  1.10$\pm$   .04 &   .47$\pm$   .03 &    .2$\pm$    .1 &   2.7$\pm$    .2 \\
  6 &  1.62$\pm$   .05 &   .40$\pm$   .02 &  -.05$\pm$   .08 &   2.6$\pm$    .1 \\
  7 &  2.11$\pm$   .05 &   .33$\pm$   .01 &  -.20$\pm$   .08 &   2.7$\pm$    .2 \\
  8 &  2.56$\pm$   .05 &   .28$\pm$   .01 &   -.2$\pm$    .1 &   2.9$\pm$    .4 \\
  9 &  2.97$\pm$   .06 &   .23$\pm$   .01 &    .0$\pm$    .2 &   2.6$\pm$    .3 \\

 \hline
 \hline
\multicolumn{5}{c}{ E$_0$ =  10000. GeV} \\
\hline
 Rad. bin & average & sigma  &skewness & kurtosis \\
\hline
  1 & -1.02$\pm$   .28 &   .98$\pm$   .11 &    .7$\pm$    .3 &   5.0$\pm$    .6 \\
  2 &  -.56$\pm$   .12 &   .73$\pm$   .08 &    .6$\pm$    .3 &   3.7$\pm$   1.1 \\
  3 &   .03$\pm$   .04 &   .62$\pm$   .05 &    .6$\pm$    .1 &   3.5$\pm$    .2 \\
  4 &   .55$\pm$   .03 &   .55$\pm$   .03 &   .38$\pm$   .06 &   3.1$\pm$    .2 \\
  5 &  1.08$\pm$   .03 &   .48$\pm$   .02 &   .24$\pm$   .08 &   2.6$\pm$    .1 \\
  6 &  1.59$\pm$   .03 &   .41$\pm$   .01 &   .01$\pm$   .05 &  2.46$\pm$   .09 \\
  7 &  2.09$\pm$   .02 &   .34$\pm$   .01 &  -.14$\pm$   .07 &   2.6$\pm$    .1 \\
  8 &  2.55$\pm$   .03 &   .28$\pm$   .01 &  -.19$\pm$   .07 &   2.7$\pm$    .1 \\
  9 &  2.96$\pm$   .03 &   .23$\pm$   .01 &    .0$\pm$    .2 &   2.9$\pm$    .4 \\

 \hline
 \hline
\multicolumn{5}{c}{ E$_0$ =  17780. GeV} \\
\hline
 Rad. bin & average & sigma  &skewness & kurtosis \\
\hline
  1 & -1.20$\pm$   .15 &   .85$\pm$   .06 &    .7$\pm$    .2 &   3.4$\pm$    .3 \\
  2 &  -.53$\pm$   .05 &   .70$\pm$   .04 &    .6$\pm$    .1 &   3.7$\pm$    .3 \\
  3 &   .02$\pm$   .03 &   .61$\pm$   .03 &   .51$\pm$   .06 &   3.4$\pm$    .1 \\
  4 &   .55$\pm$   .02 &   .53$\pm$   .02 &   .35$\pm$   .04 &   3.0$\pm$    .1 \\
  5 &  1.10$\pm$   .02 &   .46$\pm$   .01 &   .18$\pm$   .04 &  2.62$\pm$   .07 \\
  6 &  1.62$\pm$   .03 &   .40$\pm$   .01 &  -.01$\pm$   .02 &  2.58$\pm$   .06 \\
  7 &  2.12$\pm$   .03 &   .33$\pm$   .01 &  -.18$\pm$   .04 &   2.6$\pm$    .1 \\
  8 &  2.58$\pm$   .02 &   .27$\pm$   .01 &  -.16$\pm$   .08 &   2.8$\pm$    .1 \\
  9 &  3.00$\pm$   .03 &   .23$\pm$   .01 &    .0$\pm$    .1 &   3.0$\pm$    .2 \\

 \hline
 \hline
\multicolumn{5}{c}{ E$_0$ =  31620. GeV} \\
\hline
 Rad. bin & average & sigma  &skewness & kurtosis \\
\hline
  1 & -1.28$\pm$   .16 &   .89$\pm$   .03 &    .5$\pm$    .1 &   3.6$\pm$    .2 \\
  2 &  -.52$\pm$   .01 &   .68$\pm$   .02 &   .70$\pm$   .09 &   3.7$\pm$    .2 \\
  3 &   .02$\pm$   .01 &   .61$\pm$   .02 &   .54$\pm$   .06 &  3.48$\pm$   .09 \\
  4 &   .56$\pm$   .01 &   .54$\pm$   .02 &   .39$\pm$   .05 &  2.98$\pm$   .05 \\
  5 &  1.10$\pm$   .01 &   .46$\pm$   .01 &   .15$\pm$   .02 &  2.67$\pm$   .04 \\
  6 &  1.62$\pm$   .01 &   .40$\pm$   .01 &  -.03$\pm$   .01 &  2.53$\pm$   .07 \\
  7 &  2.11$\pm$   .02 &   .33$\pm$   .01 &  -.18$\pm$   .05 &  2.65$\pm$   .06 \\
  8 &  2.57$\pm$   .01 &   .27$\pm$   .01 &  -.16$\pm$   .03 &  2.72$\pm$   .07 \\
  9 &  2.99$\pm$   .02 &   .23$\pm$   .01 &   -.1$\pm$    .2 &   3.0$\pm$    .2 \\

 \hline
 \hline
\multicolumn{5}{c}{ E$_0$ =  56230. GeV} \\
\hline
 Rad. bin & average & sigma  &skewness & kurtosis \\
\hline
  1 & -1.32$\pm$   .13 &   .88$\pm$   .04 &    .4$\pm$    .2 &   3.5$\pm$    .4 \\
  2 &  -.54$\pm$   .02 &   .67$\pm$   .03 &   .58$\pm$   .06 &   3.9$\pm$    .2 \\
  3 &   .00$\pm$   .01 &   .59$\pm$   .02 &   .50$\pm$   .02 &  3.63$\pm$   .07 \\
  4 &   .56$\pm$   .01 &   .52$\pm$   .01 &   .37$\pm$   .03 &  3.17$\pm$   .04 \\
  5 &  1.11$\pm$   .01 &   .46$\pm$   .01 &   .16$\pm$   .03 &  2.73$\pm$   .05 \\
  6 &  1.64$\pm$   .02 &  -.39$\pm$   .01 &   .05$\pm$   .03 &  2.58$\pm$   .03 \\
  7 &  2.14$\pm$   .02 &   .32$\pm$   .01 &  -.19$\pm$   .03 &  2.71$\pm$   .05 \\
  8 &  2.60$\pm$   .02 &   .27$\pm$   .01 &  -.15$\pm$   .03 &   2.8$\pm$    .2 \\
  9 &  3.02$\pm$   .02 &   .22$\pm$   .01 &  -.01$\pm$   .08 &   3.0$\pm$    .2 \\

  \hline
 \hline
\multicolumn{5}{c}{ E$_0$ = 100000. GeV} \\
\hline
 Rad. bin & average & sigma  &skewness & kurtosis \\
\hline
  1 & -1.39$\pm$   .11 &   .88$\pm$   .02 &    .3$\pm$    .2 &   3.8$\pm$    .3 \\
  2 &  -.55$\pm$   .02 &   .65$\pm$   .03 &   .57$\pm$   .05 &   4.0$\pm$    .2 \\
  3 &   .01$\pm$   .01 &   .58$\pm$   .02 &   .50$\pm$   .02 &  3.56$\pm$   .07 \\
  4 &   .57$\pm$   .02 &   .51$\pm$   .01 &   .34$\pm$   .03 &  3.12$\pm$   .03 \\
  5 &  1.12$\pm$   .02 &   .45$\pm$   .01 &   .14$\pm$   .03 &  2.77$\pm$   .04 \\
  6 &  1.65$\pm$   .03 &  -.39$\pm$   .01 &   .07$\pm$   .03 &  2.58$\pm$   .04 \\
  7 &  2.15$\pm$   .02 &  -.32$\pm$   .01 &   .18$\pm$   .02 &   2.7$\pm$    .1 \\
  8 &  2.61$\pm$   .02 &   .27$\pm$   .00 &  -.15$\pm$   .03 &   2.9$\pm$    .1 \\
  9 &  3.02$\pm$   .02 &   .22$\pm$   .01 &  -.01$\pm$   .08 &   3.0$\pm$    .2 \\

 \hline
 \end{supertabular}

\clearpage

\tablefirsthead{
\\
  \multicolumn{5}{c}{Primary $\gamma$, Sec. $e^+e^-$, h$_{det}$=2000 m a.s.l.} 
\tbsp  
}

\tablehead{\hline 

  \multicolumn{5}{l}{\small\sl continued from previous page}

\\ \hline\tbsp  }

\tabletail{\hline
  \multicolumn{5}{r}{\small\sl continued on next page}
\\\hline}

\tablelasttail{\\ }

\topcaption{
Central momenta for the distribution function of 
$\log\left(\frac{t}{ns}\right)$ for secondary $e^+e^-$ produced at 2000 m a.s.l. by
a $\gamma$ primary, as obtained
  from the FLUKA stand-alone simulation}

\begin{supertabular}{ccccc}
\\ 
 \hline
\multicolumn{5}{c}{ E$_0$ =    562. GeV} \\
\hline
 Rad. bin & average & sigma  &skewness & kurtosis \\
\hline
  1 &  -.45$\pm$   .11 &  1.13$\pm$   .20 & -1.05$\pm$    .6 &  3.43$\pm$    .6\\
  2 &   .03$\pm$   .05 &   .80$\pm$   .15 &    .1$\pm$    .4 &   2.4$\pm$    .5 \\
  3 &   .35$\pm$   .24 &   .64$\pm$   .14 &    .0$\pm$    .6 &   2.4$\pm$    .6 \\
  4 &   .65$\pm$   .22 &   .55$\pm$   .15 &   -.3$\pm$    .5 &   3.0$\pm$    .5 \\
  5 &  1.09$\pm$   .18 &   .42$\pm$   .11 &   -.1$\pm$    .7 &   3.1$\pm$    .7 \\
  6 &  1.40$\pm$   .14 &   .41$\pm$   .13 &    .0$\pm$    .5 &   3.8$\pm$    .6 \\
  7 &  1.85$\pm$   .16 &   .31$\pm$   .11 &    .2$\pm$    .7 &   3.8$\pm$    .8 \\
  8 &  2.38$\pm$   .12 &   .28$\pm$   .05 &    .0$\pm$    .3 &   2.8$\pm$    .4 \\
  9 &  2.78$\pm$   .12 &   .19$\pm$   .05 &  -.52$\pm$    .3 &  2.75$\pm$    .4 \\

\hline
\hline
\multicolumn{5}{c}{E$_0$ =  1000. GeV} \\ 
\hline
 Rad. bin & average & sigma  &skewness & kurtosis \\
\hline
  1 &  -.39$\pm$   .51 &   .93$\pm$   .20 &    .0$\pm$    .6 &   2.2$\pm$    .5 \\
  2 &   .06$\pm$   .27 &   .70$\pm$   .13 &   -.2$\pm$    .5 &   2.2$\pm$    .9 \\
  3 &   .31$\pm$   .20 &   .63$\pm$   .14 &    .0$\pm$    .5 &   2.5$\pm$    .6 \\
  4 &   .73$\pm$   .24 &   .54$\pm$   .14 &   -.2$\pm$    .6 &   2.7$\pm$    .8 \\
  5 &  1.06$\pm$   .19 &   .42$\pm$   .09 &   -.1$\pm$    .5 &   3.1$\pm$    .8 \\
  6 &  1.44$\pm$   .15 &   .36$\pm$   .09 &    .1$\pm$    .6 &   3.4$\pm$   1.0 \\
  7 &  1.83$\pm$   .11 &   .30$\pm$   .07 &    .0$\pm$    .6 &   2.6$\pm$    .6 \\
  8 &  2.36$\pm$   .15 &   .28$\pm$   .06 &    .2$\pm$    .8 &   3.1$\pm$   1.0 \\
  9 &  2.83$\pm$   .15 &   .24$\pm$   .06 &   .91$\pm$    .8 &  4.10$\pm$   1.0 \\

 \hline
 \hline
\multicolumn{5}{c}{ E$_0$ =   1778. GeV} \\
\hline
 Rad. bin & average & sigma  &skewness & kurtosis \\
\hline
  1 &  -.54$\pm$   .38 &   .95$\pm$   .18 &   -.2$\pm$    .4 &   2.5$\pm$    .4 \\
  2 &   .04$\pm$   .33 &   .73$\pm$   .15 &   -.2$\pm$    .5 &   2.5$\pm$    .6 \\
  3 &   .33$\pm$   .25 &   .61$\pm$   .16 &   -.2$\pm$    .5 &   2.5$\pm$    .8 \\
  4 &   .73$\pm$   .19 &   .51$\pm$   .11 &   -.3$\pm$    .5 &   2.5$\pm$    .6 \\
  5 &  1.05$\pm$   .12 &   .41$\pm$   .07 &   -.1$\pm$    .5 &   2.7$\pm$   1.0 \\
  6 &  1.45$\pm$   .10 &   .34$\pm$   .07 &   -.1$\pm$    .6 &   3.3$\pm$   1.0 \\
  7 &  1.88$\pm$   .12 &   .29$\pm$   .07 &    .0$\pm$    .5 &   2.9$\pm$    .7 \\
  8 &  2.35$\pm$   .10 &   .27$\pm$   .06 &    .3$\pm$    .7 &   2.9$\pm$    .9 \\
  9 &  2.91$\pm$   .10 &   .26$\pm$   .06 &  -.24$\pm$    .8 &  3.03$\pm$   .9 \\

 \hline
 \hline
\multicolumn{5}{c}{ E$_0$ =   3162. GeV} \\
\hline
 Rad. bin & average & sigma  &skewness & kurtosis \\
\hline
  1 &  -.56$\pm$   .40 &   .84$\pm$   .12 &   -.1$\pm$    .4 &   2.4$\pm$    .5 \\
  2 &  -.09$\pm$   .25 &   .69$\pm$   .14 &    .1$\pm$    .4 &   2.8$\pm$    .8 \\
  3 &   .33$\pm$   .15 &   .60$\pm$   .09 &    .2$\pm$    .3 &   2.7$\pm$    .4 \\
  4 &   .73$\pm$   .11 &   .50$\pm$   .06 &    .0$\pm$    .3 &   2.5$\pm$    .5 \\
  5 &  1.07$\pm$   .09 &   .41$\pm$   .05 &    .0$\pm$    .4 &   2.8$\pm$    .6 \\
  6 &  1.44$\pm$   .07 &   .34$\pm$   .06 &    .2$\pm$    .4 &   3.0$\pm$    .6 \\
  7 &  1.90$\pm$   .08 &   .30$\pm$   .06 &    .0$\pm$    .5 &   3.5$\pm$   1.1 \\
  8 &  2.39$\pm$   .09 &   .27$\pm$   .07 &    .2$\pm$    .5 &   2.6$\pm$    .7 \\
  9 &  2.88$\pm$   .09 &   .24$\pm$   .07 &   .05$\pm$    .5 &  2.34$\pm$    .7 \\

 \hline
 \hline
\multicolumn{5}{c}{ E$_0$ =   5623. GeV} \\
\hline
 Rad. bin & average & sigma  &skewness & kurtosis \\
\hline
  1 &  -.89$\pm$   .33 &  1.03$\pm$   .19 &   -.3$\pm$    .4 &   3.0$\pm$    .9 \\
  2 &  -.14$\pm$   .20 &   .67$\pm$   .07 &    .3$\pm$    .3 &   2.7$\pm$    .4 \\
  3 &   .32$\pm$   .15 &   .58$\pm$   .07 &   -.2$\pm$    .3 &   2.5$\pm$    .3 \\
  4 &   .75$\pm$   .10 &   .48$\pm$   .05 &   -.2$\pm$    .3 &   2.6$\pm$    .4 \\
  5 &  1.11$\pm$   .08 &   .40$\pm$   .03 &   -.1$\pm$    .3 &   2.7$\pm$    .4 \\
  6 &  1.48$\pm$   .07 &   .33$\pm$   .03 &    .0$\pm$    .2 &   3.2$\pm$    .4 \\
  7 &  1.92$\pm$   .06 &   .29$\pm$   .03 &    .0$\pm$    .4 &   3.0$\pm$    .5 \\
  8 &  2.40$\pm$   .09 &   .27$\pm$   .05 &    .1$\pm$    .6 &   3.2$\pm$    .8 \\
  9 &  2.90$\pm$   .09 &   .26$\pm$   .05 &   .04$\pm$     .6&  2.26$\pm$    .8 \\

 \hline
 \hline
\multicolumn{5}{c}{ E$_0$ =  10000. GeV} \\
\hline
 Rad. bin & average & sigma  &skewness & kurtosis \\
\hline
  1 &  -.67$\pm$   .31 &   .96$\pm$   .19 &   -.3$\pm$    .5 &   3.0$\pm$    .5 \\
  2 &  -.13$\pm$   .16 &   .76$\pm$   .10 &    .0$\pm$    .3 &   2.5$\pm$    .4 \\
  3 &   .34$\pm$   .07 &   .56$\pm$   .05 &    .0$\pm$    .1 &   1.5$\pm$    .2 \\
  4 &   .74$\pm$   .04 &   .48$\pm$   .03 &    .0$\pm$    .1 &   2.7$\pm$    .4 \\
  5 &  1.08$\pm$   .05 &   .40$\pm$   .03 &    .0$\pm$    .1 &   2.5$\pm$    .2 \\
  6 &  1.44$\pm$   .04 &   .33$\pm$   .02 &    .1$\pm$    .2 &   2.9$\pm$    .3 \\
  7 &  1.90$\pm$   .04 &   .30$\pm$   .03 &   -.1$\pm$    .3 &   2.7$\pm$    .4 \\
  8 &  2.38$\pm$   .05 &   .26$\pm$   .03 &    .2$\pm$    .3 &   3.1$\pm$    .5 \\
  9 &  2.85$\pm$   .09 &   .25$\pm$   .04 &    .5$\pm$    .3 &   2.5$\pm$    .4 \\

 \hline
 \hline
\multicolumn{5}{c}{ E$_0$ =  17780. GeV} \\
\hline
 Rad. bin & average & sigma  &skewness & kurtosis \\
\hline
  1 &  -.61$\pm$   .15 &   .92$\pm$   .13 &    .1$\pm$    .2 &   2.5$\pm$    .4 \\
  2 &  -.05$\pm$   .05 &   .68$\pm$   .04 &    .1$\pm$    .2 &   2.3$\pm$    .3 \\
  3 &   .35$\pm$   .05 &   .57$\pm$   .03 &    .0$\pm$    .1 &   2.3$\pm$    .1 \\
  4 &   .73$\pm$   .03 &   .48$\pm$   .02 &   -.2$\pm$    .1 &   2.6$\pm$    .2 \\
  5 &  1.11$\pm$   .03 &   .41$\pm$   .01 &  -.06$\pm$   .10 &   2.6$\pm$    .2 \\
  6 &  1.47$\pm$   .03 &   .32$\pm$   .02 &    .0$\pm$    .2 &   3.0$\pm$    .6 \\
  7 &  1.92$\pm$   .03 &   .28$\pm$   .02 &    .2$\pm$    .2 &   3.0$\pm$    .2 \\
  8 &  2.43$\pm$   .06 &   .27$\pm$   .02 &    .2$\pm$    .2 &   3.1$\pm$    .4 \\
  9 &  2.88$\pm$   .07 &   .23$\pm$   .04 &    .1$\pm$    .3 &   2.2$\pm$    .3 \\

 \hline
 \hline
\multicolumn{5}{c}{ E$_0$ =  31620. GeV} \\
\hline
 Rad. bin & average & sigma  &skewness & kurtosis \\
\hline
  1 &  -.65$\pm$   .19 &  1.21$\pm$   .06 &   .11$\pm$   .07 &   5.0$\pm$    .1 \\
  2 &  -.05$\pm$   .04 &   .68$\pm$   .02 &    .2$\pm$    .1 &   2.2$\pm$    .2 \\
  3 &   .35$\pm$   .03 &   .56$\pm$   .01 &  -.01$\pm$   .09 &   2.1$\pm$    .1 \\
  4 &   .75$\pm$   .02 &   .49$\pm$   .02 &  -.16$\pm$   .04 &   2.4$\pm$    .1 \\
  5 &  1.10$\pm$   .01 &   .41$\pm$   .01 &   .01$\pm$   .08 &   2.9$\pm$    .1 \\
  6 &  1.47$\pm$   .02 &   .33$\pm$   .01 &   .15$\pm$   .08 &   2.9$\pm$    .2 \\
  7 &  1.92$\pm$   .01 &   .29$\pm$   .01 &    .1$\pm$    .1 &   3.1$\pm$    .3 \\
  8 &  2.39$\pm$   .02 &   .26$\pm$   .01 &    .0$\pm$    .2 &   2.7$\pm$    .3 \\
  9 &  2.88$\pm$   .05 &   .23$\pm$   .02 &    .3$\pm$    .6 &   3.0$\pm$    .9 \\

 \hline
 \hline
\multicolumn{5}{c}{ E$_0$ =  56230. GeV} \\
\hline
 Rad. bin & average & sigma  &skewness & kurtosis \\
\hline
  1 &  -.73$\pm$   .11 &   .90$\pm$   .06 &   -.1$\pm$    .1 &   2.5$\pm$    .3 \\
  2 &  -.09$\pm$   .03 &   .68$\pm$   .01 &   .12$\pm$   .07 &   2.3$\pm$    .1 \\
  3 &   .35$\pm$   .02 &   .57$\pm$   .01 &   .00$\pm$   .04 &  2.45$\pm$   .08 \\
  4 &   .75$\pm$   .01 &   .47$\pm$   .01 &  -.01$\pm$   .06 &  2.53$\pm$   .06 \\
  5 &  1.12$\pm$   .02 &   .40$\pm$   .01 &  -.04$\pm$   .06 &   2.7$\pm$    .1 \\
  6 &  1.49$\pm$   .02 &   .33$\pm$   .01 &   .04$\pm$   .06 &   2.8$\pm$    .1 \\
  7 &  1.94$\pm$   .02 &   .29$\pm$   .01 &    .1$\pm$    .1 &   3.0$\pm$    .2 \\
  8 &  2.43$\pm$   .03 &   .25$\pm$   .01 &    .1$\pm$    .1 &   2.7$\pm$    .4 \\
  9 &  2.92$\pm$   .07 &   .25$\pm$   .02 &    .2$\pm$    .3 &   3.1$\pm$    .6 \\

 \hline
 \hline
\multicolumn{5}{c}{ E$_0$ = 100000. GeV} \\
\hline
 Rad. bin & average & sigma  &skewness & kurtosis \\
\hline
  1 &  -.79$\pm$   .11 &   .91$\pm$   .01 &  -.02$\pm$   .06 &   2.6$\pm$    .2 \\
  2 &  -.08$\pm$   .02 &   .66$\pm$   .02 &   .14$\pm$   .05 &  2.34$\pm$   .06 \\
  3 &   .35$\pm$   .02 &   .56$\pm$   .02 &   .01$\pm$   .03 &  2.48$\pm$   .05 \\
  4 &   .77$\pm$   .02 &   .47$\pm$   .01 &  -.06$\pm$   .04 &  2.60$\pm$   .04 \\
  5 &  1.14$\pm$   .03 &   .40$\pm$   .01 &  -.01$\pm$   .03 &  2.69$\pm$   .04 \\
  6 &  1.51$\pm$   .03 &   .33$\pm$   .01 &   .09$\pm$   .04 &   2.8$\pm$    .2 \\
  7 &  1.96$\pm$   .02 &   .29$\pm$   .00 &   .11$\pm$   .08 &   2.9$\pm$    .2 \\
  8 &  2.45$\pm$   .03 &   .27$\pm$   .01 &    .1$\pm$    .1 &   3.1$\pm$    .2 \\
  9 &  2.93$\pm$   .03 &   .22$\pm$   .02 &    .1$\pm$    .2 &   2.9$\pm$    .4 \\

 \hline
 \end{supertabular}

\clearpage

\tablefirsthead{
\\
  \multicolumn{5}{c}{Primary protons, Sec. $\gamma$, h$_{det}$=2000 m a.s.l.} 
\tbsp  
}

\tablehead{\hline 

  \multicolumn{5}{l}{\small\sl continued from previous page}

\\ \hline\tbsp  }

\tabletail{\hline
  \multicolumn{5}{r}{\small\sl continued on next page}
\\\hline}

\tablelasttail{\\ }

\topcaption{
Central momenta for the distribution function of 
$\log\left(\frac{t}{ns}\right)$ for secondary $\gamma$ produced at 2000 m a.s.l. by
a proton primary, as obtained
  from the FLUKA stand-alone simulation}

\begin{supertabular}{ccccc}
\\ 
 \hline
\multicolumn{5}{c}{ E$_0$ =    562. GeV} \\
\hline
 Rad. bin & average & sigma  &skewness & kurtosis \\
\hline
  1 &   .20$\pm$   .40 &   .59$\pm$   .14 &  1.12$\pm$    .7 &  3.65$\pm$   1.4 \\
  2 &   .38$\pm$   .34 &   .63$\pm$   .13 &    .8$\pm$    .6 &   3.3$\pm$   1.4 \\
  3 &   .55$\pm$   .44 &   .64$\pm$   .18 &   1.1$\pm$    .9 &   3.9$\pm$   2.8 \\
  4 &   .62$\pm$   .37 &   .64$\pm$   .15 &    .8$\pm$    .6 &   2.6$\pm$   1.1 \\
  5 &   .99$\pm$   .28 &   .58$\pm$   .14 &    .2$\pm$    .6 &   2.7$\pm$   1.1 \\
  6 &  1.44$\pm$   .25 &   .50$\pm$   .13 &    .1$\pm$    .7 &   2.6$\pm$   2.3 \\
  7 &  1.92$\pm$   .21 &   .44$\pm$   .12 &   -.1$\pm$    .8 &   2.9$\pm$   2.4 \\
  8 &  2.36$\pm$   .25 &   .41$\pm$   .14 &   -.1$\pm$    .9 &   3.1$\pm$   2.8 \\
  9 &  2.88$\pm$   .34 &   .56$\pm$   .19 &    .6$\pm$    .9 &   5.6$\pm$   2.2 \\

\hline
\hline
\multicolumn{5}{c}{E$_0$ =  1000. GeV} \\ 
\hline
 Rad. bin & average & sigma  &skewness & kurtosis \\
\hline
  1 &  -.25$\pm$   .30 &   .59$\pm$   .19 &   1.3$\pm$    .9 &   4.5$\pm$   3.1 \\
  2 &  -.35$\pm$   .45 &   .74$\pm$   .14 &    .9$\pm$    .5 &   3.2$\pm$   1.4 \\
  3 &  -.04$\pm$   .33 &   .64$\pm$   .15 &   1.1$\pm$    .5 &   3.2$\pm$   1.2 \\
  4 &   .46$\pm$   .36 &   .62$\pm$   .11 &    .4$\pm$    .5 &   2.6$\pm$   1.3 \\
  5 &  1.04$\pm$   .27 &   .52$\pm$   .13 &    .1$\pm$    .5 &   2.9$\pm$   1.6 \\
  6 &  1.51$\pm$   .22 &   .48$\pm$   .11 &   -.2$\pm$    .5 &   2.8$\pm$   1.4 \\
  7 &  1.99$\pm$   .21 &   .44$\pm$   .11 &   -.5$\pm$    .7 &   3.7$\pm$   2.7 \\
  8 &  2.45$\pm$   .24 &   .38$\pm$   .12 &   -.2$\pm$    .9 &   3.1$\pm$   3.1 \\
  9 &  2.91$\pm$   .32 &   .54$\pm$   .17 &    .5$\pm$    .9 &   5.5$\pm$   2.4 \\

 \hline
 \hline
\multicolumn{5}{c}{ E$_0$ =   1778. GeV} \\
\hline
 Rad. bin & average & sigma  &skewness & kurtosis \\
\hline
  1 &   .04$\pm$   .24 &   .66$\pm$   .07 &    .4$\pm$    .9 &   3.0$\pm$   2.4 \\
  2 &  -.10$\pm$   .40 &   .82$\pm$   .18 &    .4$\pm$    .6 &   2.8$\pm$    .8 \\
  3 &   .12$\pm$   .42 &   .72$\pm$   .17 &    .8$\pm$    .6 &   2.7$\pm$    .9 \\
  4 &   .45$\pm$   .32 &   .64$\pm$   .11 &    .4$\pm$    .5 &   2.7$\pm$    .9 \\
  5 &   .97$\pm$   .23 &   .54$\pm$   .09 &    .1$\pm$    .5 &   2.5$\pm$   2.2 \\
  6 &  1.49$\pm$   .17 &   .47$\pm$   .11 &    .0$\pm$    .5 &   2.5$\pm$   1.9 \\
  7 &  1.99$\pm$   .14 &   .40$\pm$   .08 &   -.2$\pm$    .8 &   2.7$\pm$   2.9 \\
  8 &  2.44$\pm$   .19 &   .39$\pm$   .11 &   -.2$\pm$    .8 &   3.5$\pm$   2.8 \\
  9 &  2.93$\pm$   .24 &   .50$\pm$   .13 &    .5$\pm$    .9 &   6.0$\pm$   2.6 \\

 \hline
 \hline
\multicolumn{5}{c}{ E$_0$ =   3162. GeV} \\
\hline
 Rad. bin & average & sigma  &skewness & kurtosis \\
\hline
  1 &  -.60$\pm$   .44 &   .78$\pm$   .18 &   1.2$\pm$    .6 &   4.2$\pm$   1.5 \\
  2 &  -.53$\pm$   .41 &   .78$\pm$   .13 &    .6$\pm$    .6 &   2.9$\pm$   1.4 \\
  3 &  -.09$\pm$   .42 &   .71$\pm$   .13 &    .3$\pm$    .6 &   2.9$\pm$    .8 \\
  4 &   .48$\pm$   .26 &   .59$\pm$   .12 &    .2$\pm$    .4 &   3.1$\pm$    .9 \\
  5 &  1.02$\pm$   .18 &   .53$\pm$   .10 &    .0$\pm$    .4 &   2.7$\pm$    .8 \\
  6 &  1.53$\pm$   .13 &   .45$\pm$   .07 &   -.1$\pm$    .4 &   2.6$\pm$   1.7 \\
  7 &  2.03$\pm$   .13 &   .40$\pm$   .06 &   -.3$\pm$    .6 &   3.1$\pm$   2.5 \\
  8 &  2.47$\pm$   .17 &   .37$\pm$   .07 &   -.3$\pm$    .8 &   3.2$\pm$   2.9 \\
  9 &  2.95$\pm$   .25 &   .50$\pm$   .14 &    .5$\pm$    .7 &   5.5$\pm$   2.1 \\

 \hline
 \hline
\multicolumn{5}{c}{ E$_0$ =   5623. GeV} \\
\hline
 Rad. bin & average & sigma  &skewness & kurtosis \\
\hline
  1 &  -.72$\pm$   .48 &   .82$\pm$   .25 &    .3$\pm$   1.0 &   2.7$\pm$   4.2 \\
  2 &  -.55$\pm$   .63 &   .83$\pm$   .17 &    .8$\pm$    .6 &   3.1$\pm$    .6 \\
  3 &   .03$\pm$   .52 &   .65$\pm$   .19 &    .5$\pm$    .5 &   3.2$\pm$  30.3 \\
  4 &   .47$\pm$   .23 &   .61$\pm$   .09 &    .3$\pm$    .4 &   2.8$\pm$   2.2 \\
  5 &  1.01$\pm$   .12 &   .54$\pm$   .10 &    .1$\pm$    .3 &   2.8$\pm$    .6 \\
  6 &  1.47$\pm$   .13 &   .55$\pm$   .10 &   -.5$\pm$    .4 &   4.1$\pm$   1.4 \\
  7 &  2.03$\pm$   .10 &   .40$\pm$   .08 &   -.3$\pm$    .4 &   3.1$\pm$   1.5 \\
  8 &  2.46$\pm$   .15 &   .37$\pm$   .09 &   -.3$\pm$    .7 &   3.3$\pm$   2.7 \\
  9 &  2.92$\pm$   .24 &   .49$\pm$   .12 &    .3$\pm$    .7 &   5.7$\pm$   2.0 \\

 \hline
 \hline
\multicolumn{5}{c}{ E$_0$ =  10000. GeV} \\
\hline
 Rad. bin & average & sigma  &skewness & kurtosis \\
\hline
  1 &  -.61$\pm$   .39 &   .89$\pm$   .20 &    .2$\pm$    .5 &   2.4$\pm$    .7 \\
  2 &  -.46$\pm$   .36 &   .81$\pm$   .16 &    .5$\pm$    .4 &   2.8$\pm$    .8 \\
  3 &  -.05$\pm$   .24 &   .71$\pm$   .08 &    .4$\pm$    .3 &   3.1$\pm$    .5 \\
  4 &   .49$\pm$   .12 &   .60$\pm$   .07 &    .3$\pm$    .2 &   2.9$\pm$    .3 \\
  5 &  1.02$\pm$   .07 &   .52$\pm$   .05 &    .1$\pm$    .2 &   2.8$\pm$    .7 \\
  6 &  1.54$\pm$   .10 &   .45$\pm$   .07 &   -.1$\pm$    .2 &   2.8$\pm$    .9 \\
  7 &  2.03$\pm$   .07 &  -.39$\pm$   .04 &    .2$\pm$    .3 &   2.8$\pm$   1.5 \\
  8 &  2.49$\pm$   .08 &   .34$\pm$   .05 &   -.2$\pm$    .6 &   3.2$\pm$   2.4 \\
  9 &  2.96$\pm$   .11 &   .43$\pm$   .07 &    .5$\pm$    .6 &   5.3$\pm$   1.8 \\

 \hline
 \hline
\multicolumn{5}{c}{ E$_0$ =  17780. GeV} \\
\hline
 Rad. bin & average & sigma  &skewness & kurtosis \\
\hline
  1 &  -.62$\pm$   .41 &  1.13$\pm$   .13 &    .4$\pm$    .3 &   5.8$\pm$    .6 \\
  2 &  -.47$\pm$   .33 &   .75$\pm$   .04 &    .5$\pm$    .3 &   2.7$\pm$    .6 \\
  3 &  -.12$\pm$   .17 &   .72$\pm$   .07 &    .2$\pm$    .2 &   3.4$\pm$    .4 \\
  4 &   .48$\pm$   .10 &   .58$\pm$   .05 &    .2$\pm$    .1 &   2.8$\pm$    .3 \\
  5 &  1.06$\pm$   .03 &   .49$\pm$   .02 &   .13$\pm$   .06 &   2.7$\pm$    .3 \\
  6 &  1.57$\pm$   .07 &   .43$\pm$   .02 &   -.2$\pm$    .2 &   2.7$\pm$    .5 \\
  7 &  2.06$\pm$   .05 &   .37$\pm$   .02 &   -.2$\pm$    .3 &   2.8$\pm$   1.0 \\
  8 &  2.51$\pm$   .06 &   .33$\pm$   .03 &   -.3$\pm$    .4 &   3.4$\pm$   1.4 \\
  9 &  2.97$\pm$   .03 &   .44$\pm$   .03 &    .3$\pm$    .5 &   5.9$\pm$   1.2 \\

 \hline
 \hline
\multicolumn{5}{c}{ E$_0$ =  31620. GeV} \\
\hline
 Rad. bin & average & sigma  &skewness & kurtosis \\
\hline
  1 & -1.34$\pm$   .52 &   .92$\pm$   .07 &    .4$\pm$    .4 &   3.3$\pm$    .5 \\
  2 &  -.56$\pm$   .24 &   .74$\pm$   .08 &    .5$\pm$    .2 &   3.6$\pm$    .5 \\
  3 &  -.02$\pm$   .12 &   .64$\pm$   .06 &    .3$\pm$    .2 &   3.3$\pm$    .3 \\
  4 &   .53$\pm$   .04 &   .56$\pm$   .04 &   .25$\pm$   .07 &   3.0$\pm$    .2 \\
  5 &  1.07$\pm$   .05 &   .49$\pm$   .03 &    .0$\pm$    .1 &   2.7$\pm$    .3 \\
  6 &  1.59$\pm$   .05 &  -.43$\pm$   .02 &    .1$\pm$    .2 &   2.7$\pm$    .4 \\
  7 &  2.09$\pm$   .05 &   .37$\pm$   .03 &   -.3$\pm$    .3 &   3.0$\pm$   1.1 \\
  8 &  2.53$\pm$   .06 &   .33$\pm$   .04 &   -.3$\pm$    .5 &   3.4$\pm$   1.2 \\
  9 &  3.00$\pm$   .08 &   .42$\pm$   .03 &    .7$\pm$    .4 &   5.5$\pm$    .8 \\

 \hline
 \hline
\multicolumn{5}{c}{ E$_0$ =  56230. GeV} \\
\hline
 Rad. bin & average & sigma  &skewness & kurtosis \\
\hline
  1 & -1.29$\pm$   .33 &   .88$\pm$   .05 &    .5$\pm$    .2 &   3.4$\pm$    .3 \\
  2 &  -.58$\pm$   .05 &   .70$\pm$   .05 &   .51$\pm$   .09 &   3.5$\pm$    .2 \\
  3 &  -.03$\pm$   .03 &   .62$\pm$   .04 &   .46$\pm$   .03 &   3.3$\pm$    .2 \\
  4 &   .52$\pm$   .04 &   .55$\pm$   .02 &   .27$\pm$   .06 &   3.0$\pm$    .1 \\
  5 &  1.08$\pm$   .03 &   .48$\pm$   .02 &   .12$\pm$   .05 &   2.7$\pm$    .1 \\
  6 &  1.61$\pm$   .03 &   .41$\pm$   .01 &   -.1$\pm$    .1 &   2.7$\pm$    .3 \\
  7 &  2.10$\pm$   .04 &   .36$\pm$   .02 &   -.3$\pm$    .1 &   2.9$\pm$    .6 \\
  8 &  2.56$\pm$   .04 &   .31$\pm$   .02 &   -.2$\pm$    .4 &   3.2$\pm$    .8 \\
  9 &  3.05$\pm$   .05 &   .43$\pm$   .02 &    .7$\pm$    .2 &   5.8$\pm$    .5 \\

 \hline
 \hline
\multicolumn{5}{c}{ E$_0$ = 100000. GeV} \\
\hline
 Rad. bin & average & sigma  &skewness & kurtosis \\
\hline
  1 & -1.25$\pm$   .11 &   .88$\pm$   .05 &    .5$\pm$    .1 &   3.3$\pm$    .2 \\
  2 &  -.58$\pm$   .08 &   .71$\pm$   .06 &    .5$\pm$    .1 &   3.7$\pm$    .2 \\
  3 &  -.03$\pm$   .04 &   .64$\pm$   .03 &   .39$\pm$   .09 &   3.4$\pm$    .2 \\
  4 &   .52$\pm$   .02 &   .55$\pm$   .02 &   .28$\pm$   .05 &  3.04$\pm$   .07 \\
  5 &  1.08$\pm$   .03 &   .48$\pm$   .02 &   .08$\pm$   .02 &   2.8$\pm$    .1 \\
  6 &  1.61$\pm$   .03 &  -.41$\pm$   .01 &   .11$\pm$   .05 &   2.7$\pm$    .2 \\
  7 &  2.10$\pm$   .02 &   .35$\pm$   .01 &   -.3$\pm$    .1 &   2.9$\pm$    .6 \\
  8 &  2.56$\pm$   .03 &   .31$\pm$   .02 &   -.2$\pm$    .2 &   3.2$\pm$    .6 \\
  9 &  3.02$\pm$   .04 &   .40$\pm$   .02 &    .6$\pm$    .2 &   5.1$\pm$    .6 \\

 \hline
 \end{supertabular}

\clearpage

\tablefirsthead{
\\
  \multicolumn{5}{c}{Primary protons, Sec. $e^+e^-$, h$_{det}$=2000 m a.s.l.} 
\tbsp  
}

\tablehead{\hline 

  \multicolumn{5}{l}{\small\sl continued from previous page}

\\ \hline\tbsp  }

\tabletail{\hline
  \multicolumn{5}{r}{\small\sl continued on next page}
\\\hline}

\tablelasttail{\\ }

\topcaption{
Central momenta for the distribution function of 
$\log\left(\frac{t}{ns}\right)$ for secondary $e^+e^-$ produced at 2000 m a.s.l. by
a proton primary, as obtained
  from the FLUKA stand-alone simulation}

\begin{supertabular}{ccccc}
\\ 
 \hline
\multicolumn{5}{c}{ E$_0$ =    562. GeV} \\
\hline
 Rad. bin & average & sigma  &skewness & kurtosis \\
\hline
  1 &   .26$\pm$   .30 &   .43$\pm$   .10 &   .37$\pm$    .7 &  1.90$\pm$    .10 \\
  2 &   .49$\pm$   .27 &   .63$\pm$   .11 &   .85$\pm$    .6 &  3.74$\pm$    .8 \\
  3 &   .57$\pm$   .27 &   .59$\pm$   .11 &    .2$\pm$    .6 &   2.2$\pm$    .8 \\
  4 &   .66$\pm$   .26 &   .54$\pm$   .11 &    .2$\pm$    .7 &   2.1$\pm$   1.4 \\
  5 &   .93$\pm$   .28 &   .50$\pm$   .12 &   -.2$\pm$    .8 &   3.4$\pm$   2.4 \\
  6 &  1.30$\pm$   .23 &   .43$\pm$   .12 &    .0$\pm$    .7 &   2.9$\pm$   1.7 \\
  7 &  1.72$\pm$   .28 &   .43$\pm$   .19 &    .1$\pm$   1.0 &   3.5$\pm$   1.8 \\
  8 &  2.23$\pm$   .36 &   .44$\pm$   .20 &    .4$\pm$   1.0 &   3.5$\pm$   1.5 \\
  9 &  2.74$\pm$   .34 &   .60$\pm$   .27 &    .3$\pm$   1.0 &   5.8$\pm$   5.8 \\

\hline
\hline
\multicolumn{5}{c}{E$_0$ =  1000. GeV} \\ 
\hline
 Rad. bin & average & sigma  &skewness & kurtosis \\
\hline
  1 &   .17$\pm$   .29 &   .69$\pm$   .19 &    .2$\pm$    .5 &   2.2$\pm$    .6 \\
  2 &  -.03$\pm$   .36 &   .79$\pm$   .15 &    .2$\pm$    .5 &   2.1$\pm$    .8 \\
  3 &   .17$\pm$   .25 &   .68$\pm$   .11 &    .2$\pm$    .5 &   2.3$\pm$    .6 \\
  4 &   .60$\pm$   .17 &   .54$\pm$   .11 &   -.2$\pm$    .6 &   2.4$\pm$    .8 \\
  5 &  1.02$\pm$   .26 &   .47$\pm$   .12 &   -.1$\pm$    .6 &   3.3$\pm$   1.0 \\
  6 &  1.34$\pm$   .24 &   .43$\pm$   .12 &   -.1$\pm$    .8 &   3.6$\pm$   3.1 \\
  7 &  1.81$\pm$   .28 &   .58$\pm$   .17 &   -.2$\pm$   1.1 &   5.8$\pm$   3.7 \\
  8 &  2.32$\pm$   .34 &   .56$\pm$   .19 &   1.0$\pm$   1.0 &   5.9$\pm$   2.1 \\
  9 &  2.77$\pm$   .37 &   .50$\pm$   .18 &    .2$\pm$    .9 &   3.2$\pm$   1.9 \\

 \hline
 \hline
\multicolumn{5}{c}{ E$_0$ =   1778. GeV} \\
\hline
 Rad. bin & average & sigma  &skewness & kurtosis \\
\hline
  1 &   .42$\pm$   .40 &   .62$\pm$   .20 &   .54$\pm$    .5 &  2.50$\pm$    .4 \\
  2 &   .15$\pm$   .40 &   .78$\pm$   .19 &   -.2$\pm$    .4 &   2.3$\pm$    .3 \\
  3 &   .28$\pm$   .21 &   .69$\pm$   .13 &    .0$\pm$    .5 &   2.4$\pm$    .7 \\
  4 &   .56$\pm$   .18 &   .56$\pm$   .13 &   -.1$\pm$    .6 &   2.0$\pm$    .8 \\
  5 &   .95$\pm$   .18 &   .46$\pm$   .11 &    .0$\pm$    .7 &   2.8$\pm$   1.6 \\
  6 &  1.33$\pm$   .19 &   .41$\pm$   .10 &    .2$\pm$    .7 &   3.2$\pm$   1.9 \\
  7 &  1.81$\pm$   .20 &   .36$\pm$   .11 &    .3$\pm$    .8 &   3.1$\pm$   1.9 \\
  8 &  2.31$\pm$   .29 &   .51$\pm$   .16 &    .2$\pm$   1.0 &   4.8$\pm$   2.3 \\
  9 &  2.86$\pm$   .24 &   .55$\pm$   .20 &   1.1$\pm$    .9 &   5.1$\pm$   1.8 \\

 \hline
 \hline
\multicolumn{5}{c}{ E$_0$ =   3162. GeV} \\
\hline
 Rad. bin & average & sigma  &skewness & kurtosis \\
\hline
  1 &  -.32$\pm$   .30 &   .77$\pm$   .20 &    .1$\pm$    .5 &   2.2$\pm$    .8 \\
  2 &  -.14$\pm$   .31 &   .80$\pm$   .14 &   -.2$\pm$    .4 &   2.7$\pm$    .9 \\
  3 &   .20$\pm$   .23 &   .67$\pm$   .11 &   -.2$\pm$    .4 &   2.5$\pm$    .5 \\
  4 &   .63$\pm$   .18 &   .53$\pm$   .09 &   -.1$\pm$    .4 &   2.8$\pm$   1.0 \\
  5 &  1.01$\pm$   .18 &   .47$\pm$   .11 &   -.2$\pm$    .4 &   2.9$\pm$    .6 \\
  6 &  1.36$\pm$   .15 &   .40$\pm$   .09 &   -.2$\pm$    .8 &   3.3$\pm$   3.8 \\
  7 &  1.82$\pm$   .19 &   .37$\pm$   .13 &   -.1$\pm$   1.0 &   3.1$\pm$   3.6 \\
  8 &  2.32$\pm$   .26 &   .42$\pm$   .18 &    .3$\pm$    .8 &   3.7$\pm$   2.1 \\
  9 &  2.88$\pm$   .37 &   .58$\pm$   .20 &    .0$\pm$    .7 &   5.5$\pm$   1.5 \\

 \hline
 \hline
\multicolumn{5}{c}{ E$_0$ =   5623. GeV} \\
\hline
 Rad. bin & average & sigma  &skewness & kurtosis \\
\hline
  1 &  -.31$\pm$   .27 &   .85$\pm$   .20 &   -.1$\pm$    .6 &   2.4$\pm$   1.0 \\
  2 &  -.20$\pm$   .46 &   .86$\pm$   .16 &   -.1$\pm$    .4 &   2.5$\pm$    .6 \\
  3 &   .31$\pm$   .24 &   .63$\pm$   .14 &   -.1$\pm$    .6 &   2.4$\pm$   1.1 \\
  4 &   .59$\pm$   .26 &   .55$\pm$   .11 &   -.4$\pm$    .4 &   2.9$\pm$    .5 \\
  5 &  1.00$\pm$   .14 &   .47$\pm$   .10 &   -.2$\pm$    .5 &   2.8$\pm$   2.1 \\
  6 &  1.29$\pm$   .15 &   .50$\pm$   .11 &   -.5$\pm$    .8 &   4.7$\pm$   2.4 \\
  7 &  1.83$\pm$   .13 &   .40$\pm$   .11 &    .0$\pm$    .5 &   3.8$\pm$   1.5 \\
  8 &  2.34$\pm$   .20 &   .41$\pm$   .16 &    .4$\pm$    .9 &   4.1$\pm$   2.1 \\
  9 &  2.83$\pm$   .35 &   .65$\pm$   .19 &    .5$\pm$    .9 &   4.3$\pm$   1.6 \\

 \hline
 \hline
\multicolumn{5}{c}{ E$_0$ =  10000. GeV} \\
\hline
 Rad. bin & average & sigma  &skewness & kurtosis \\
\hline
  1 &  -.16$\pm$   .32 &   .83$\pm$   .19 &   -.3$\pm$    .4 &   2.7$\pm$    .7 \\
  2 &  -.15$\pm$   .28 &   .79$\pm$   .15 &   -.2$\pm$    .4 &   1.8$\pm$    .5 \\
  3 &   .23$\pm$   .22 &   .68$\pm$   .10 &   -.3$\pm$    .3 &   2.3$\pm$    .6 \\
  4 &   .63$\pm$   .12 &   .55$\pm$   .08 &   -.2$\pm$    .3 &   2.7$\pm$    .3 \\
  5 &  1.01$\pm$   .10 &   .46$\pm$   .06 &   -.2$\pm$    .5 &   2.9$\pm$   1.4 \\
  6 &  1.38$\pm$   .13 &   .39$\pm$   .07 &   -.1$\pm$    .8 &   3.2$\pm$   4.4 \\
  7 &  1.81$\pm$   .10 &   .37$\pm$   .07 &    .1$\pm$    .9 &   3.3$\pm$   4.1 \\
  8 &  2.33$\pm$   .14 &   .38$\pm$   .09 &    .2$\pm$    .9 &   3.4$\pm$   2.6 \\
  9 &  2.88$\pm$   .22 &   .56$\pm$   .15 &    .5$\pm$    .7 &   5.7$\pm$   1.6 \\

 \hline
 \hline
\multicolumn{5}{c}{ E$_0$ =  17780. GeV} \\
\hline
 Rad. bin & average & sigma  &skewness & kurtosis \\
\hline
  1 &  -.22$\pm$   .42 &   .82$\pm$   .14 &   -.1$\pm$    .3 &   2.7$\pm$    .3 \\
  2 &  -.12$\pm$   .26 &   .78$\pm$   .08 &   -.2$\pm$    .5 &   3.4$\pm$   1.1 \\
  3 &   .21$\pm$   .09 &   .67$\pm$   .06 &   -.1$\pm$    .2 &   2.9$\pm$    .4 \\
  4 &   .63$\pm$   .07 &   .53$\pm$   .05 &  -.24$\pm$   .10 &   2.7$\pm$    .3 \\
  5 &  1.06$\pm$   .04 &   .43$\pm$   .04 &   -.1$\pm$    .2 &   2.8$\pm$    .5 \\
  6 &  1.40$\pm$   .06 &   .37$\pm$   .03 &    .0$\pm$    .5 &   2.9$\pm$   3.5 \\
  7 &  1.85$\pm$   .06 &   .34$\pm$   .03 &    .1$\pm$   1.3 &   3.4$\pm$   5.8 \\
  8 &  2.34$\pm$   .14 &   .38$\pm$   .09 &    .1$\pm$    .8 &   3.8$\pm$   2.1 \\
  9 &  2.93$\pm$   .13 &   .50$\pm$   .08 &   1.1$\pm$    .5 &   4.5$\pm$   1.5 \\

 \hline
 \hline
\multicolumn{5}{c}{ E$_0$ =  31620. GeV} \\
\hline
 Rad. bin & average & sigma  &skewness & kurtosis \\
\hline
  1 &  -.78$\pm$   .43 &   .91$\pm$   .13 &    .0$\pm$    .3 &   2.0$\pm$    .5 \\
  2 &  -.14$\pm$   .21 &   .71$\pm$   .08 &   -.1$\pm$    .2 &   2.6$\pm$    .3 \\
  3 &   .29$\pm$   .12 &   .62$\pm$   .06 &   -.2$\pm$    .3 &   2.8$\pm$    .8 \\
  4 &   .70$\pm$   .05 &   .52$\pm$   .03 &  -.14$\pm$   .10 &   2.7$\pm$    .3 \\
  5 &  1.07$\pm$   .06 &   .44$\pm$   .03 &   -.2$\pm$    .1 &   3.0$\pm$    .6 \\
  6 &  1.43$\pm$   .06 &   .37$\pm$   .03 &   -.1$\pm$    .3 &   3.0$\pm$   1.2 \\
  7 &  1.87$\pm$   .05 &   .35$\pm$   .03 &    .0$\pm$    .9 &   3.2$\pm$   3.4 \\
  8 &  2.38$\pm$   .08 &   .38$\pm$   .07 &    .3$\pm$    .8 &   4.2$\pm$   2.3 \\
  9 &  2.88$\pm$   .15 &   .47$\pm$   .08 &    .5$\pm$    .5 &   5.5$\pm$   1.0 \\

 \hline
 \hline
\multicolumn{5}{c}{ E$_0$ =  56230. GeV} \\
\hline
 Rad. bin & average & sigma  &skewness & kurtosis \\
\hline
  1 &  -.73$\pm$   .23 &   .90$\pm$   .06 &    .1$\pm$    .2 &   1.9$\pm$    .3 \\
  2 &  -.17$\pm$   .02 &   .70$\pm$   .04 &    .0$\pm$    .1 &   2.5$\pm$    .3 \\
  3 &   .29$\pm$   .05 &   .61$\pm$   .03 &  -.02$\pm$   .07 &  2.29$\pm$   .09 \\
  4 &   .69$\pm$   .04 &   .51$\pm$   .02 &  -.15$\pm$   .08 &   2.4$\pm$    .4 \\
  5 &  1.08$\pm$   .03 &   .43$\pm$   .02 &   -.1$\pm$    .1 &   2.7$\pm$    .4 \\
  6 &  1.45$\pm$   .03 &   .36$\pm$   .02 &    .0$\pm$    .3 &   3.0$\pm$   1.9 \\
  7 &  1.92$\pm$   .04 &   .34$\pm$   .02 &    .2$\pm$    .3 &   3.3$\pm$   2.1 \\
  8 &  2.44$\pm$   .06 &   .44$\pm$   .04 &    .3$\pm$    .2 &   4.9$\pm$   1.0 \\
  9 &  2.98$\pm$   .09 &   .51$\pm$   .04 &    .5$\pm$    .3 &   6.1$\pm$    .5 \\

 \hline
 \hline
\multicolumn{5}{c}{ E$_0$ = 100000. GeV} \\
\hline
 Rad. bin & average & sigma  &skewness & kurtosis \\
\hline
  1 &  -.72$\pm$   .14 &   .90$\pm$   .03 &    .1$\pm$    .2 &   1.9$\pm$    .2 \\
  2 &  -.14$\pm$   .08 &   .70$\pm$   .06 &    .0$\pm$    .1 &   2.3$\pm$    .1 \\
  3 &   .29$\pm$   .04 &   .61$\pm$   .03 &  -.13$\pm$   .07 &   2.6$\pm$    .1 \\
  4 &   .69$\pm$   .02 &   .52$\pm$   .02 &  -.14$\pm$   .04 &  2.68$\pm$   .09 \\
  5 &  1.08$\pm$   .03 &   .43$\pm$   .01 &  -.08$\pm$   .06 &   2.8$\pm$    .2 \\
  6 &  1.45$\pm$   .03 &   .36$\pm$   .01 &    .0$\pm$    .2 &   3.0$\pm$   1.9 \\
  7 &  1.90$\pm$   .03 &   .34$\pm$   .02 &    .0$\pm$    .4 &   3.2$\pm$   2.1 \\
  8 &  2.42$\pm$   .04 &   .37$\pm$   .03 &    .4$\pm$    .2 &   4.0$\pm$   1.2 \\
  9 &  2.96$\pm$   .07 &   .51$\pm$   .05 &    .6$\pm$    .3 &   5.7$\pm$   1.1 \\

 \hline
 \end{supertabular}

 \end{center}

\end{document}